\documentclass[twocolumn,prb,aps,showpacs,superscriptaddress,floatfix]{revtex4}
\usepackage{array}
\usepackage{dcolumn}
\usepackage{longtable}
\usepackage{amsfonts}
\usepackage{amssymb}
\usepackage{amsmath}
\usepackage{hyperref}
\usepackage{color}
\usepackage[dvips]{graphicx}
\usepackage[american]{babel}
\usepackage{bm}
\usepackage{subfigure}
\newcommand{\cop}{\hat{c}}
\newcommand{\nop}{\hat{n}}
\newcommand{\Nop}{\hat{N}}
\newcommand{\one}{\hat{1}}
\newcommand{\Xop}{\hat{X}}
\newcommand{\sop}{\hat{s}}
\newcommand{\Sop}{\hat{S}}
\newcommand{\Lop}{\hat{L}}
\newcommand{\upr}{\uparrow}
\newcommand{\dwr}{\downarrow}
\newcommand{\mycolumnwidth}{0.92\columnwidth}
\newcommand{\halfcolumnwidth}{0.46\columnwidth}
\newcommand{\brcolumnwidth}{0.72\columnwidth}
\begin{document}
%


%
\author{L.\ Tincani}
\email[]{leonildo.tincani@physik.uni-marburg.de}
\affiliation{Fachbereich Physik, 
	     Philipps Universit{\"a}t Marburg, 
	     D-35032 Marburg, 
	     Germany}
\author{R.M.\ Noack}
\affiliation{Fachbereich Physik, 
	     Philipps Universit{\"a}t Marburg, 
	     D-35032 Marburg, 
             Germany}
\author{D.\ Baeriswyl}
\affiliation{D{\'e}partement de Physique, 
             Universit{\'e} de Fribourg, 
             CH-1700 Fribourg, 
             Switzerland}
%


%
\title{Critical properties of the band-insulator-to-Mott-insulator transition 
in the strong-coupling limit of the ionic Hubbard model}
\date{January 14, 2008}
%

%
%
%
%


%
\begin{abstract}
We investigate the neutral-to-ionic insulator-insulator transition 
in one-dimensional materials by treating a strong-coupling effective
model based on the ionic Hubbard model using the density-matrix
renormalization group and finite-size scaling. 
The effective model, formulated in a spin-one representation, 
contains a single parameter.
We carry out an extensive finite-size scaling analysis of the relevant
gaps and susceptibilities to characterize the two zero-temperature
transitions. 
We find that the transition from the ionic band-insulating phase  to an
intermediate spontaneously dimerized phase is Ising, and the
transition from the dimerized phase to the Mott-insulating phase is
Kosterlitz-Thouless, in agreement with the field-theory-based
predictions. 
\end{abstract}
\maketitle
%

\begin{figure}[htb]
  \centering
  \includegraphics[width=\mycolumnwidth,height=\brcolumnwidth]{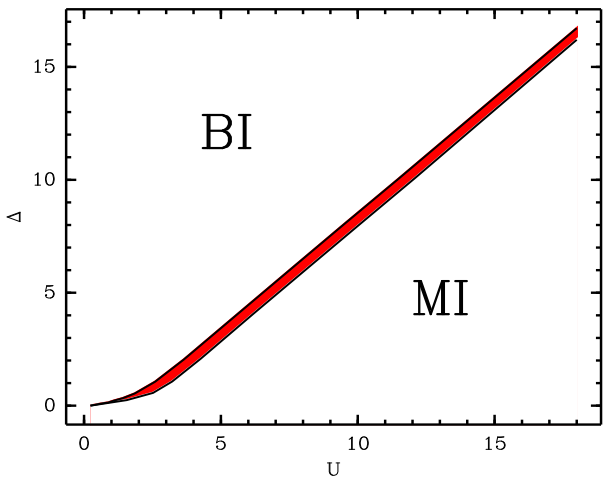}
  \caption{(Color online) Ground-state phase diagram of the ionic Hubbard model.
    Location of the phase boundaries is approximate, but drawn to
    scale according to values from Refs.~{\onlinecite{Torio_2001}} and
    {\onlinecite{Manmana_2004}}.
    The (red) shaded intermediate region designates a spontaneously dimerized
    insulating phase (SDI).}
  \label{fig:iHmPD}
\end{figure}

%
Electrons in solids are subject to both a single-particle potential
and the Coulomb interaction.
A wealth of interesting phenomena can occur when the form of the
single-particle potential deviates from that of the ideal crystal due
to, for example, structural transitions, lattice vibrations, or
defects or impurities.
A simple Hamiltonian that incorporates the combined effects of
interactions and reduced translational symmetry in a particularly
transparent manner is the ionic Hubbard model (IHM), in which the
single-particle energy alternates between neighboring sites.
This model was introduced by Nagaosa and Takimoto
\cite{Nagaosa_1986a,Nagaosa_1986b,Nagaosa_1986c}
to describe the neutral-ionic transition observed by Torrance 
{\em et al.}\ in mixed-stack organic charge-transfer compounds.~\cite{Torrance_1981}
In a mixed stack of donor (D) and acceptor (A) molecules, the neutral
phase corresponds to a uniform and neutral distribution of charge,
${\rm D^0A^0D^0A^0}$,
and the ionic phase to an alternation of positive 
and negative charges, ${\rm D^+A^-D^+A^-}$.
The insulating behavior in the neutral phase originates from the
Coulomb interaction between electrons, i.e., the Mott mechanism,
whereas the ionic phase is essentially a band insulator.
Recently, the neutral-ionic transition has been observed in organic
charge-transfer compounds close to zero temperature, 
motivating interest in it as a pure quantum phase
transition.~\cite{Horiuchi_2003}
A different class of quasi-one-dimensional materials in which a
similar charge disproportionation occurs is that of the
halogen-bridged transition-metal complexes, whose structure is formed
by a backbone of of alternating metal and halogen atoms.~\cite{Gammel_1992}
In these MX-chain compounds 
(or in the related MMX materials~\cite{Yamamoto_2001}), 
a spontaneous breaking of the translational symmetry occurs 
due to the dimerization of the halogen sublattice, XMX--M--XMX--M.
The differing distances of the halogen ions from the neighboring metal
ions give rise to a two-fold alternation in the energy of the $d$ levels.
The Hamiltonian of the ionic Hubbard model can be grouped into three terms, 
a one-dimensional nearest-neighbor hopping term with matrix element $t$, 
an on-site Coulomb repulsion of strength $U$, 
and an ionic alternating potential of depth $\Delta$,
\begin{equation}
\hat{H} = 
\hat{H}_{t} + \hat{H}_{U} + \hat{H}_{\Delta} 
\, , 
\label{equ:Hamiltonian}
\end{equation}
with
\begin{equation}
\hat{H}_{t} = 
t \sum_{i=1, \sigma}^{L-1} 
\left(
\cop_{i \sigma}^\dag \cop_{i+1 \sigma} + \cop_{i+1 \sigma}^\dag \cop_{i \sigma} 
\right)
\; , 
\label{equ:Hhopping}
\end{equation}
\begin{equation}
\hat{H}_{U} = 
\frac{U}{2} \sum_{i=1, \sigma}^{L}
\nop_{i \sigma} \nop_{i -\sigma} 
\; ,
\label{equ:HCoulomn}
\end{equation}
and
\begin{equation}
\hat{H}_{\Delta} = 
\frac{\Delta}{2} \sum_{i=1, \sigma}^{L}
\left( -1 \right)^{i} \nop_{i \sigma} 
\; .
\label{equ:Hionic}
\end{equation}
Here $\cop_{i\sigma}^\dag$ ($\cop_{i\sigma}$) 
are the usual creation (annihilation) operators on site $i$ 
for an electron of spin $\sigma$ 
and $\nop_{i\sigma} = \cop_{i\sigma}^\dag$ $\cop_{i\sigma}$.
Without the ionic potential, $\Delta=0$, 
the model reduces to the one-dimensional Hubbard model, 
whose behavior is well understood.~\cite{Essler_2005}
Although the overall physics described by the ionic Hubbard model is
now fairly well known, 
many details of the transition are still unclear.
The general behavior in the ground state is summarized in the
schematic ground-state phase diagram shown in Fig.~\ref{fig:iHmPD}. 
When $\Delta \gtrsim U$, the system is a band insulator (BI)
and has both a charge and spin gap.
When $\Delta \lesssim U$, the system is a critically antiferromagnetic
Mott insulator (MI) with a charge gap and gapless spin excitations.
These two phases are separated by two continuous phase-transition
lines within which there is a spontaneously dimerized insulating
phase (SDI) of width of order $t$, i.e., a phase with both spin and
charge gaps as well as with long-range bond dimer order.
%

%
%

%
In order to understand the origin of the phases, let us first examine
what happens in the atomic limit, 
$t=0$, which can be easily treated.
For $U>\Delta$ and at half filling, 
there is no double occupancy in the ground state, 
which consists of a series of singly occupied sites
with energy $\pm \Delta/2$ so that the entire system has energy $E=0$.
For $U<\Delta$, double occupancy is favorable, 
and the ground state consists of doubly occupied sites
at energy $U-\Delta$ alternating with empty sites, 
so that the energy of the system is $L(U-\Delta)/2$.
At $(U-\Delta) \rightarrow 0$, 
a level crossing of two configurations occurs 
and there is a first-order transition at $U=\Delta$.
Turning on the hopping term leads to more subtle behavior 
in the vicinity of the transition.
In the noninteracting limit, $U=0$, 
the Hamiltonian is diagonal in momentum space.
It follows that the ionic term, $\Delta$, opens a charge and a spin gap,
and the two gaps have the same value.
Correspondingly, spin-spin and charge-charge correlations decay
exponentially.
The scenario does not change 
with the inclusion of a weak interaction $\hat{H}_U$;
the electrons tend to doubly occupy sites with lower potential,
and the system remains a band insulator.
In the large-$U$ limit, 
the double occupancy can be treated perturbatively
and the low-energy physics of the IHM is described 
by an effective spin one-half Heisenberg model.~\cite{Nagaosa_1986a,Kampf_2003,Kakashvili_2004}
It is important to note that this effective model
restores translational invariance, 
and that the charge and spin sectors are completely separated.
The system has gapless spin excitations
and critical spin-spin correlations,
while the charge gap, in contrast, 
scales as $U$ for large $U$.
This description is robust for a wide range of parameters
in the strong coupling limit,
but fails close to the transition line 
because perturbation theory breaks down in the critical regime.~\cite{Nagaosa_1986a}
In fact, there are analytical and numerical indications
that show that higher-order spin excitations mix 
into the charge degrees of freedom everywhere 
in the MI phase.~\cite{Resta_1995,Torio_2001,Manmana_2004}
A few years ago, 
Fabrizio, Gogolin and Nersesyan proposed a new, 
interesting scenario based on field-theoretical arguments.~\cite{Fabrizio_1999}
They argued that two quantum phase transitions occur, 
an Ising transition between the band insulator 
and an intermediate spontaneously dimerized phase, 
followed by (for increasing $U/\Delta$) a Kosterlitz-Thouless transition
(KT) between the dimerized phase and the Mott insulator. 
This scenario is based on an argument in which the transition
is approached, on the one hand, from the MI limit and,
on the other hand, from the BI limit.
The authors consider the weak-coupling case,
$(U, \Delta) << t$, and use standard bosonization.
The Hamiltonian then consists of three parts,
a first term depending only on charge degrees of freedom,
a second term involving only spin degrees of freedom and a
third term, proportional to $\Delta$, which couples charge 
and spin degrees of freedom.
Starting from the MI phase ($U$ dominating) with a charge gap
but no spin gap, one can integrate out the charge degrees of freedom.
This leads to a sine-Gordon model for the spin degrees of freedom
with a positive coupling for $U> U_{c_2}$. The coupling term
turns negative for $U< U_{c_2}$, and therefore $U_{c_2}$ corresponds
to a KT transition point. A spin gap opens for $U<U_{c_2}$ and is
attributed to a spontaneously dimerized insulating phase (SDI).
Starting from the BI phase ($\Delta$ dominating), which exhibits 
both a charge and a spin gap, Fabrizio, Gogolin and Nersesyan calculate
spin and bond-order susceptibilities using perturbation theory.
A critical value $U_{c_1}$ is found where the bond-order susceptibility
diverges, while the spin susceptibility remains finite. 
Thus, $U_{c_1}$ must be in a region with a finite spin gap, and 
it follows that $U_{c_1} < U_{c_2}$.
Close to $U_{c_1}$ it is argued that the spin degrees of freedom
can be considered to be frozen. This yields a double sine-Gordon Hamiltonian
for the charge degrees of freedom, which is known to undergo 
a quantum phase transition of an Ising type.~\cite{Delfino_1998}
The order parameter of this transition is the bond order operator,
which confirms that the intermediate region, $U_{c_1} < U < U_{c_2}$,
is in a SDI phase.~\cite{Fabrizio_2000}
At least one transition
has been found in all numerical work 
\cite{Gidopoulos_2000,Wilkens_2001,Torio_2001,Kampf_2003,Lou_2003,Manmana_2004,Otsuka_2005,Legeza_2005,Legeza_2006}
published after Ref.~\onlinecite{Fabrizio_1999}, although,
for the most part, the critical behavior was not characterized.
The critical exponents were calculated in Ref.\ \onlinecite{Manmana_2004}, 
but were found to deviate from the expected two-dimensional-Ising values.
However, even confirming that there is a second transition has been a quite difficult task.
The two transitions turn out to be very close to one another and, 
since the transition to the Mott insulator 
is expected to be a KT transition, 
it is very difficult to find and characterize using
finite-size-scaling studies.~\cite{Manmana_2004} 
For these reasons, 
studying an effective model characterizing 
the region of the transition and the intermediate phase is useful.
Another very important subtlety is 
how to map the gaps from the field-theoretical model 
onto the original lattice model.
In the ionic Hubbard model, the charge gap, the one-particle gap, 
and the spin gap all behave differently at the transitions.  
The one-particle gap is related to the charge and spin gaps 
but is fundamentally different because it involves a change of the
particle number, 
while the charge and spin gaps are spectral gaps of excitations 
into the charge and spin sectors, respectively, only.
One way of locating critical points is to examine the smallest energy gap,
i.e., the mass gap, as a function of the tuning parameters. 
The critical point is then the point at which 
the gap vanishes in the thermodynamic limit.
The remainder of this paper is organized as follows.
In Sec.\ \ref{sec:emodel}, 
an effective spin-one model for the
transition is derived via a strong-coupling treatment.
In Sec.\ \ref{sec:nmcq},
the numerical method used to study the model is described.
In Sec.\ \ref{sec:bisdi} and \ref{sec:sdimi} we report the analysis of
the band-insulator-to-spontaneously-dimerized insulator 
and the spontaneously-dimerized-to-Mott insulator transitions, 
respectively.
%


%
\section{\label{sec:emodel} Effective model}
\subsection{\label{subsec:dem} Derivation of the effective Hamiltonian}
In order to investigate the critical behavior 
of the ionic Hubbard model at half filling, 
we derive an effective model, 
formulated in terms of spin-one operators, 
valid for $\left( U, \Delta \right) >> t$. 
In this limit, 
the doubly occupied state on the even sites 
(with on-site potential $\Delta/2$)
and the unoccupied state on the odd sites can be projected out.
At half filling, 
a double occupancy on an even site is necessarily associated 
with a completely unoccupied odd site, 
and has a cost in energy of $U + \Delta$.
This procedure is a second-order strong-coupling expansion 
with parameter $t/(U,\Delta)$ analogous to that used to derive the
$t$-$J$ model from the Hubbard model.
In fact, 
the resulting model can equivalently be formulated 
in terms of $t$-$J$ operators rather than spin-one operators;
we feel that the latter formulation is more intuitive 
for the half-filled system.~\cite{Aligia_2005,Tincani_2008}
The physical meaning of the spin-one states is as follows: 
the $S_z=\pm 1$ state corresponds to a singly occupied site 
with a spin-$\frac12$ electron with spin up or down, 
while the $S_z=0$ state corresponds to an unoccupied site 
on the even sites and a doubly occupied site on the odd sites.
The mapping of the states of the ionic Hubbard model 
to those of the effective spin-one model is summarized in
Table~\ref{table:mapping}.
\begingroup
\squeezetable
\begin{table}[htbp]
 \caption{Mapping between the single-site basis states 
  of the ionic Hubbard model $\{ |0\rangle , |\uparrow\rangle ,
  |\downarrow\rangle , |d\rangle\}$ with $|d\rangle$ denoting the
  doubly occupied state, and those of the effective spin-one
  model $\{ | S^z\rangle \}$.}
   \centering
  \vspace{0.5cm}
  \begin{tabular}{cccccccccc}
    \hline
    \hline \\
    \ & &  $-\Delta/2$  & & \quad \quad  \quad & \quad \quad \quad & & $+\Delta/2$ & & \ \\
    \\
    \hline
    \\
    \ & $ |\ 0 \ \rangle $  & $ \rightarrow $ &   excluded          & \ & \ & $ |\ 0 \ \rangle $  & $ \rightarrow $ & $ |\ 0 \ \rangle $ & \ \\
    \ & $ |\upr \ \rangle $ & $ \rightarrow $ & $ |\ 1 \ \rangle $  & \ & \ & $ |\upr \ \rangle $ & $ \rightarrow $ & $ |\ 1 \ \rangle $ & \ \\
    \ & $ |\dwr \ \rangle $ & $ \rightarrow $ & $ |\ -1 \ \rangle $ & \ & \ & $ |\dwr \ \rangle $ & $ \rightarrow $ & $ |\ -1\ \rangle $ & \ \\
    \ & $ |\ d \ \rangle $  & $ \rightarrow $ & $ |\ 0 \ \rangle $  & \ & \ & $ |\ d \ \rangle $  & $ \rightarrow $ &   excluded & \  \\
    \\   
    \hline
    \hline
  \end{tabular}
  \label{table:mapping}
\end{table}
\endgroup
As we shall see, 
conservation of particle number leads to a spin exchange process 
for the spin-one operators that is more restricted 
than the Heisenberg exchange.
Given the mapping of states described above, 
the effective Hamiltonian can most easily be derived
by first expressing the original Hamiltonian 
in terms of transition operators between the fermionic states (Hubbard
operators), then
projecting out the states as outlined above, and subsequently writing the
Hamiltonian in the reduced state space in terms of spin transition
operators. 
Finally, the transition operators in spin space can be rewritten in
terms of spin-one operators.~\cite{Fazekas_1999,Haley_1972}
A detailed derivation is given in the appendix.~\ref{app:modder}
The Hamiltonian for the effective spin-one model can 
thus be expressed in terms of the usual spin-one operators,
yielding $\hat{H}^e = \hat{H}^e_t + \hat{H}^e_\varepsilon$, 
with the exchange term
\begin{eqnarray}
\hat{H}_{t}^{e} = 
\frac{t}{2} & \displaystyle \sum \limits_{i=1}^L &
\left[ 
\left( \Sop^{+}_i\Sop^{-}_{i+1} 
+ \Sop^{-}_i \Sop^{+}_{i+1} \right) \Sop^{z}_{i+1}
\right. \nonumber \\
& & 
\quad
- \left.
\Sop^{z}_{i} \left( \Sop^{+}_i \Sop^{-}_{i+1} 
+ \Sop^{-}_{i} \Sop^{+}_{i+1} \right) 
\right] 
\, 
\label{equ:HeffectiveKin}
\end{eqnarray}
and the interaction term governed 
by the single parameter $\varepsilon = U - \Delta$,
\begin{equation}
\hat{H}^{e}_{\varepsilon} = 
- \frac{\varepsilon}{2} \sum_{i=1}^{L} 
\left [ \left( \Sop_i^z \right)^2 - 1 \right ]
\; .
\label{equ:HeffectivePot}
\end{equation}
Note that it is immediately clear from the effective model 
that the relevant interaction parameter is $\varepsilon = U-\Delta$.
For $t=0$, it is clear that there should be a transition at 
$U \sim \Delta$ because the sign of the $\hat{H}^{e}_{\varepsilon}$ term
changes. 
For $\varepsilon >> t$, the on-site $S^z=0$ state is strongly suppressed
so that the remaining degrees of freedom, $S^z=1$ and $S^z=-1$,
correspond to the localized spin-$\frac12$ degrees of freedom 
of the MI phase of the original model.
For $\varepsilon \to -\infty$, the $S^z=\pm 1$ local states 
are suppressed, leading to a ground state that is a simple product 
of local $S^z=0$ states, which maps to the band insulator.
However, the nature of the transition(s) 
and possible intermediate phases for finite $t$ 
still needs to be determined.
In particular, 
it is important to investigate whether 
the behavior in the vicinity of $\varepsilon = 0$ agrees 
with previous numerical results for 
the ionic Hubbard model,~\cite{Manmana_2004,Kampf_2003,Legeza_2006}
as well as with field-theoretical treatments.~\cite{Fabrizio_1999}
Note that the derivation of the effective model can easily 
be extended to include additional interaction terms 
that do not break the symmetries of the original model, 
such as a next-nearest-neighbor Coulomb repulsion.
In a similar context, a related effective model was developed 
some time ago in Ref.~\onlinecite{Horovitz_1987}.
\subsection{\label{subsec:observables} Observables }
Since the formulation of the effective model in terms of spin-one operators 
is a notational convenience rather than physical, 
we are interested in studying observables of the original model.
Therefore, 
it is necessary to translate the observables of the IHM into the language of the spin-one model.
The local spin operators map as 
(small letters: IHM, capital letters: effective model)
\begin{displaymath}
\begin{array}{ccl}
\displaystyle \sop_i^z    & \displaystyle \rightarrow & \displaystyle \frac12 \Sop_i^z \,, \\
& & \\
\displaystyle \sop_i^{\pm} & \displaystyle \rightarrow & \displaystyle \frac12 \left( \Sop_i^{\pm} \right)^2 \, ,\\
& & \\
\displaystyle \sop_i^2    & \displaystyle \rightarrow & \displaystyle \frac34 \left(\Sop_i^z \right)^2 \,,
\end{array}
\label{equar:MapObsOne}
\end{displaymath}
the local charge operators as
\begin{displaymath}
\displaystyle \nop_i \rightarrow \left\{
\begin{array}{rl}
        \left( \Sop_i^z \right)^2 & \ \text{i \ = \ even } \\
& \\
2 \, -  \left( \Sop_i^z \right)^2 & \ \text{i \ = \ odd } \,\, ,
\end{array}
\right. 
\end{displaymath}
and total spin and charge operators as
\begin{eqnarray}
 \sop^z &  \rightarrow  &  \frac12 \Sop^z \, ,\nonumber\\[0.2cm]
 \sop^2 &  \rightarrow  &  \frac12 \Sop^z \left( \frac12 \Sop^z + 1 \right) 
+ \frac14 \sum \limits_{i,j=1}^{L} \left( \Sop^-_i \right)^2 \left( \Sop^+_j \right)^2 \, ,\nonumber\\[0.2cm]
 \Nop &  \rightarrow &  L + \sum \limits_{i=1}^{L} \left( -1 \right)^i \left( \Sop_i^z \right)^2 \, .
\label{equar:MapObsTwo}
\end{eqnarray}

As we can see, 
conservation of $s^z$ in the IHM leads 
to conservation of $S^z$ in the effective model, 
with the spin scaled by a factor of one half.
However, 
conservation of the total spin 
in the IHM does not lead to conservation 
of total spin for the effective model, 
which is not $SU(2)$-invariant. 
In Table~\ref{table:mapobs} we show the mapping 
of the most important quantities from the original 
ionic Hubbard model to the effective spin-one model. 
\begingroup
\squeezetable
\begin{table*}[htbp]
  \caption{Mapping of relevant physical quantities to the effective spin one model.}
  \vspace{0.5cm}
  \centering
  \begin{tabular}{clcclcclc}
    \hline
    \hline
    \ & & \ & \ & & \ & \ & & \ \\
    \ & Quantity & \ & \ & Ionic Hubbard Model  & \ & \ & Effective Spin One Model & \ \\
    \ & & \ & \ & & \ & \ & & \ \\
    \hline
    \ & & \ & \ & & \ & \ & & \ \\
    \ & Ionicity               & \ & \ & $ \displaystyle I = \frac{2}{L} \sum \limits_{i=1}^{L} \left( -1 \right)^{i} \langle \nop_i \rangle$ & \ & \ & $ \displaystyle I = 2 - \frac{2}{L} \sum \limits_{i=1}^{L} \left \langle \left( \Sop_i^z \right)^2 \right \rangle$ & \ \\
    \ & & \ & \ & & \ & \ & & \ \\
    \ & & \ & \ & & \ & \ & & \ \\
    \ & Polarization           & \ & \ & $\displaystyle P_{e} = \frac{1}{L} \sum \limits_{i=1}^{L} x_i \langle \nop_i \rangle$ & \ & \ & $ \displaystyle P_e = \frac{1}{L} \sum \limits_{i=1}^L \left( -1 \right)^i x_i \left \langle \left( \Sop_i^z \right)^2 \right \rangle - \frac12$ & \ \\
    \ & & \ & \ & & \ & \ & & \ \\
    \ & Bond Order Parameter   & \ & \ & $\displaystyle D = \frac{1}{L-1} \sum_{i=1}^{L-1} (-1)^i \langle \cop_i^\dag \cop_{i+1} + \cop_{i+1}^\dag \cop_i \rangle $ & \ & \ &  $\displaystyle D = \frac{1}{L-1} \sum_{i=1}^L (-1)^i \left[ \left \langle \left( \Sop^{+}_i \Sop^{-}_{i+1} + \Sop^{-}_i \Sop^{+}_{i+1} \right) \Sop^{z}_{i+1} \right. \right. $ & \ \\
    \ & & \ & \ & & \ & \ & \qquad \qquad \qquad $ \left. \left. \displaystyle -  \Sop^{z}_{i} \left( \Sop^{+}_i \Sop^{-}_{i+1} + \Sop^{-}_{i}\Sop^{+}_{i+1}\right) \right \rangle \right ] $ & \ \\
    \ & & \ & \ & & \ & \ & & \ \\
    \ & AFM Order               & \ & \ & $\displaystyle A = \frac{1}{L} \sum_i^{L} (-1)^i \langle \sop^z_i \rangle$ & \ & \ & $\displaystyle A = \frac{1}{2L} \sum_i^{L} (-1)^i \langle \Sop^z_i \rangle$ & \ \\
    \ & & \ & \ & & \ & \ & & \ \\
    \hline
    \hline
  \end{tabular}
  \label{table:mapobs}
\end{table*}
\endgroup

\subsection{Symmetries}
One relevant characteristic of the effective model is the extent 
to which the symmetries of the original model are preserved or modified.
The interaction term $\hat{H}_{\varepsilon}^{e}$ is local,
translationally invariant, and depends only on $(S^z)^2$, 
in contrast to the on-site part of the IHM Hamiltonian in 
Eq.~(\ref{equ:Hamiltonian}).
The apparently greater translational symmetry of the effective model is
a consequence 
of the reduction of state space in transforming to the effective model.
Note that this is only true at half filling: the quantity 
$\langle \hat{N}\rangle -L$ (see Eq.~\ref{equar:MapObsTwo})
is conserved and breaks translational symmetry except at half filling,
where it is zero.
(Note that the {\it interpretation} of the $S^z=0$ state is not translationally invariant.)
Since the spin-exchange term has the same symmetries 
as the hopping term in the IHM, the remaining symmetries 
of the original model are preserved in the effective model.
Conserved quantities in the original model, 
such as the total z-component of the spin, $s_z$, 
the total spin, $s$, and the number of particles, $N$,
are still conserved in the effective model, but have different meanings.
%


%
\section{\label{sec:nmcq} Numerical Method}
We have investigated the effective model by performing
density-matrix renormalization group (DMRG) calculations for different
system sizes,
from $L=200$ up to $600$ sites, with open boundary conditions (OBC)
and an even number of sites.~\cite{White_1992,Schollwock_2005}
For small chains, boundary effects can be large, 
depending on the correlation length.
Thus, in order to minimize any dispersion due to the edges,~\cite{Fabrizio_1995} 
Friedel oscillations,~\cite{Bedurftig_1998} and odd-even effects,~\cite{Hotta_2006}
we analyze systems of at least $200$ sites.
In order to achieve sufficient accuracy,
at least 5 sweeps must be performed,
with up to 1280 states retained in the last sweep.
The maximum system size that can be accurately treated is 
then approximately 600 sites.~\cite{Legeza_1996}
The maximum discarded weight of the density matrix
for the effective model is always less than $10^{-8}$,
and is typically zero to within the numerical precision 
far from the critical points.~\cite{Legeza_2003b}
In order to calculate ground-state properties, we target the ground
state in the $S^z=0$ sector; we target  
both the ground state and the first excited state in the $S^z=0$
sector to calculate the `exciton' gap of the original IHM; 
and the lowest states in the $S^z=1$ and $S^z=2$ sectors are needed to
calculate the charge and spin gaps, respectively, of the
IHM.~\cite{Manmana_2004}
We have repeated the same calculations
using the dynamic block-state selection (DBSS) approach,
fixing the threshold of maximum quantum information loss 
to $\chi = 10^{-6}$ at each step.~\cite{Legeza_2003,Legeza_2003b}
For instance, $m\approx500$ basis states are enough 
to correctly describe the ground-state wave function 
of a system with $500$ sites for $\varepsilon = 1.23$.
However, as we increase $\varepsilon$ 
the number of states required increases,
for example to 
$m \approx 900$ states for $\varepsilon=2$.
For ground states of other symmetry sectors,
e.g., the lowest triplet excitation,
this number can sometimes be larger
when the excited state is delocalized,
despite the fact that its Fock subspace is smaller.
Nevertheless, since we are interested 
in only the energy of these states 
and since measurements are carried out only on the absolute ground state, 
keeping of the order of a thousand states is usually sufficient.
As the aim of the effective model is to describe
the strong-coupling limit of the IHM when $( U,\Delta )>> t$,
we have compared results from the effective model to DMRG results for the IHM 
for $U \simeq \Delta = 20 t$.~\cite{Manmana_2004}
All the quantities that we measure: gaps, ionicity, 
bond order parameter and polarization, 
are in agreement to within a few percent.
%


%
\section{\label{sec:bisdi} BI to SDI transition}
In this section
we study the first transition
between the band-insulator phase
and the spontaneously dimerized phase.
We have tuned the interaction coupling $\varepsilon$ 
starting from zero,
where the system behaves like a band insulator,
increasing it until the first transition point $\varepsilon_{c_1}$ is reached.
In order to locate the transition point,
we have studied the behavior of the singlet and triplet gaps
and of the bond order parameter.
The two gaps go to zero in the thermodynamic limit 
at the transition point and subsequently reopen.
The value of the bond order parameter,
which measures the system's dimerization,
changes from zero to a finite value across the transition.
The existence of such a transition has been
extensively discussed for the IHM.~\cite{Manmana_2004,Legeza_2006}
Therefore, 
we have focused on the characterization of the transition
by evaluating its critical exponents explicitly. 
The Hamiltonian of the effective model is new 
and is not evidently related to any known classical model.
Therefore, 
we must first determine the value of the dynamic critical exponent $z$
in order to carry out finite-size scaling.
Subsequently,
we extract the correlation length exponent $\nu$
from the divergence of the mass gap.
Finally,
the thermodynamic exponents $\beta$, $\alpha$ and $\gamma$,
which are all related to the free energy density,
are obtained by analyzing the divergence of the bond order parameter, 
the specific heat, and the bond-order susceptibility, respectively.
\subsection{Dynamic critical exponent $z$}
\begin{figure}[htb]
  \centering
  \subfigure[]{
    \includegraphics[width=\halfcolumnwidth,height=\brcolumnwidth]{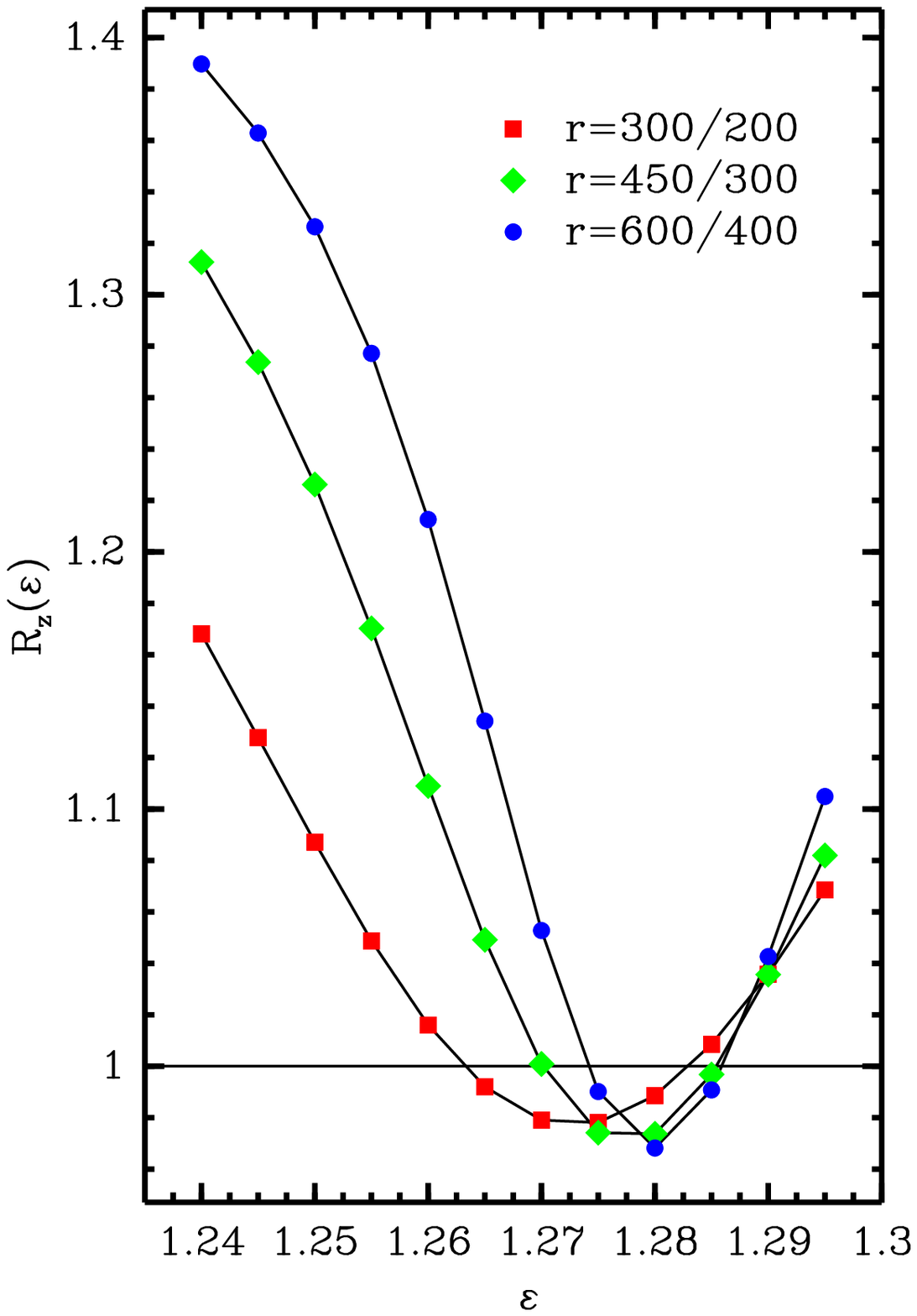}
    \label{fig:ratiozA}}
  \subfigure[]{
    \includegraphics[width=\halfcolumnwidth,height=\brcolumnwidth]{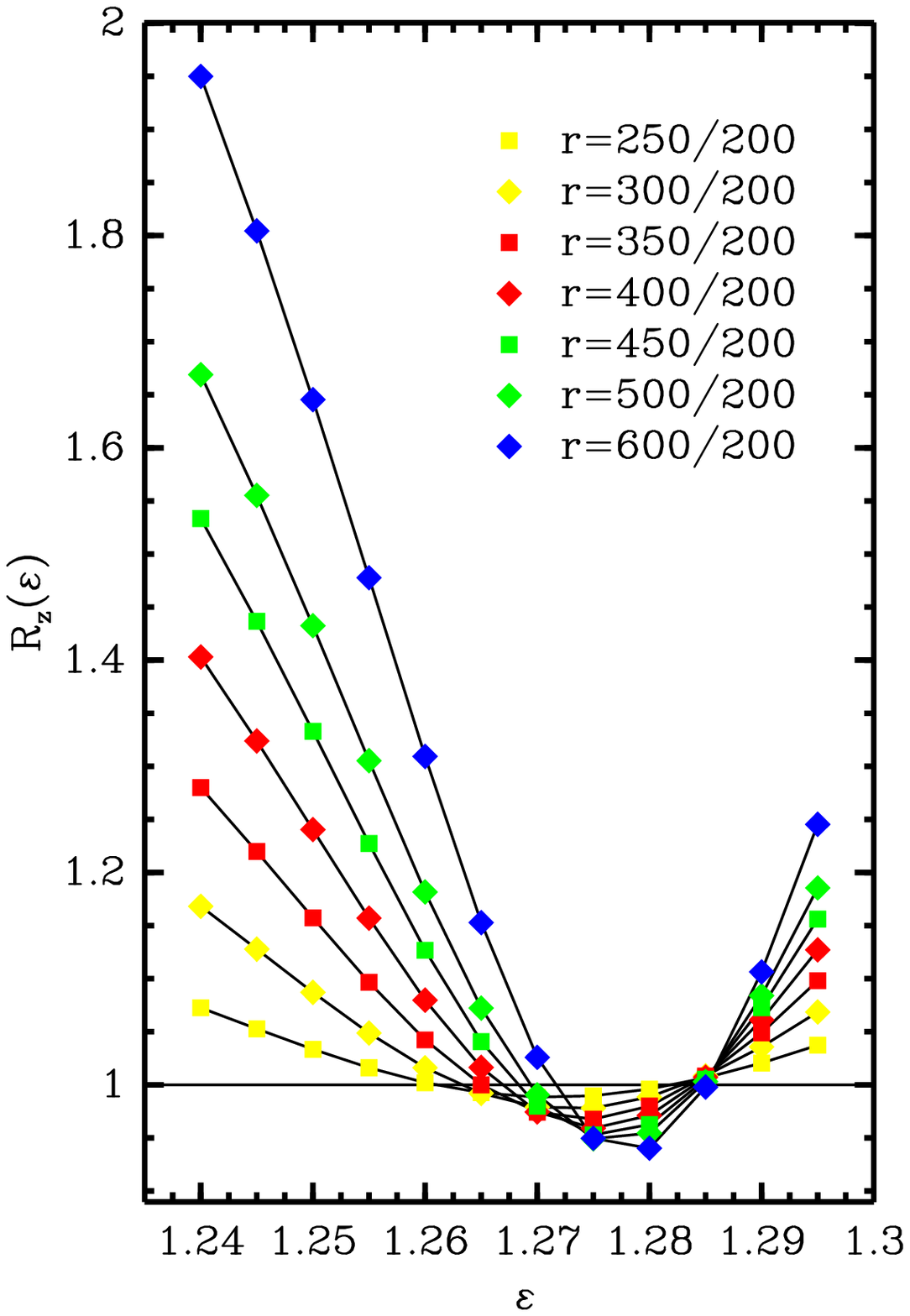}
    \label{fig:ratiozB}}
  \caption[Different system-size gap ratios as function of the coupling $\varepsilon$]{
    (Color online) 
    Mass gap ratio $R_{z}$ as function of the coupling $\varepsilon$ for
    \subref{fig:ratiozA} various system sizes $N$ and $M$ and the same
    ratio $r=1.5$, and 
    \subref{fig:ratiozB} mass gap ratio for different $r$ scaled by the $M=200$ gap.}
  \label{fig:ratioz}
\end{figure}

For a quantum system related to a classical model by the transfer matrix,
the dynamic critical exponent plays the role of an extra dimension,
i.e., $z=1$.
In general, 
space and time correlations can be coupled, 
and the value of $z$ can be different from one.
Therefore,
a determination of $z$ is required to obtain 
and interpret all the remaining critical exponents.
First, 
we identify the mass gap
\begin{equation}
F \left( \varepsilon, L \right) = 
E_{1} \left(\varepsilon , L \right) - E_{0} \left( \varepsilon , L \right)
\, ,
\label{equ:masgap}
\end{equation}
which is the gap that scales to zero most quickly close to the critical
point.~\cite{Kogut_1979}
This gap is proportional to $\xi^{-z}$, where the correlation length
$\xi$ is limited by the system size $L$.
Consequently, the ratio
\begin{equation}
R_{z}(\varepsilon, N,M) = 
\frac{F \left( \varepsilon, N \right) }{F \left(\varepsilon, M\right)} \frac{N}{M}
\label{eq:genmassgapratio}
\end{equation}
of the mass gaps for different system sizes behaves 
as $R_{z}(\varepsilon_{c_1}, N, M) \sim \left( N/ M \right)^{1-z}$
for $N,M >> 1$, and thus depends only on the ratio of system sizes
$r\equiv N/M$.~\cite{Barber_1983}
In Fig.~\ref{fig:ratiozA}, 
we show that all the gap ratios with a particular
$r$ ($r=1.5$ in the figure) cross each other at the same point, which
is near $R_{z}=1$.
The behavior is similar for other values of $r$; we have examined 
$r=1.2$, 1.25, 1.33, and 2.
In Fig.~\ref{fig:ratiozB}, one can see that 
curves with different $r$, scaled by the $M=200$ gap,
cross $R_{z}=1$ at the same point.
Thus, 
it is clear that all curves cross each other 
at approximately the same value of $\varepsilon$, $\varepsilon \approx 1.3$,
where $R_{z}(N,M) \approx 1$, consistent with $z=1$.~\cite{Glaus_1984}
In order to carry out the scaling analysis of the critical coupling
and other critical exponents, we take $z=1$ in the following subsections.
%
%
%
%

%
\subsection{Correlation length exponent $\nu$}

\begin{figure}[htb]
  \centering
 \subfigure[]{
   \includegraphics[width=\halfcolumnwidth,height=\brcolumnwidth]{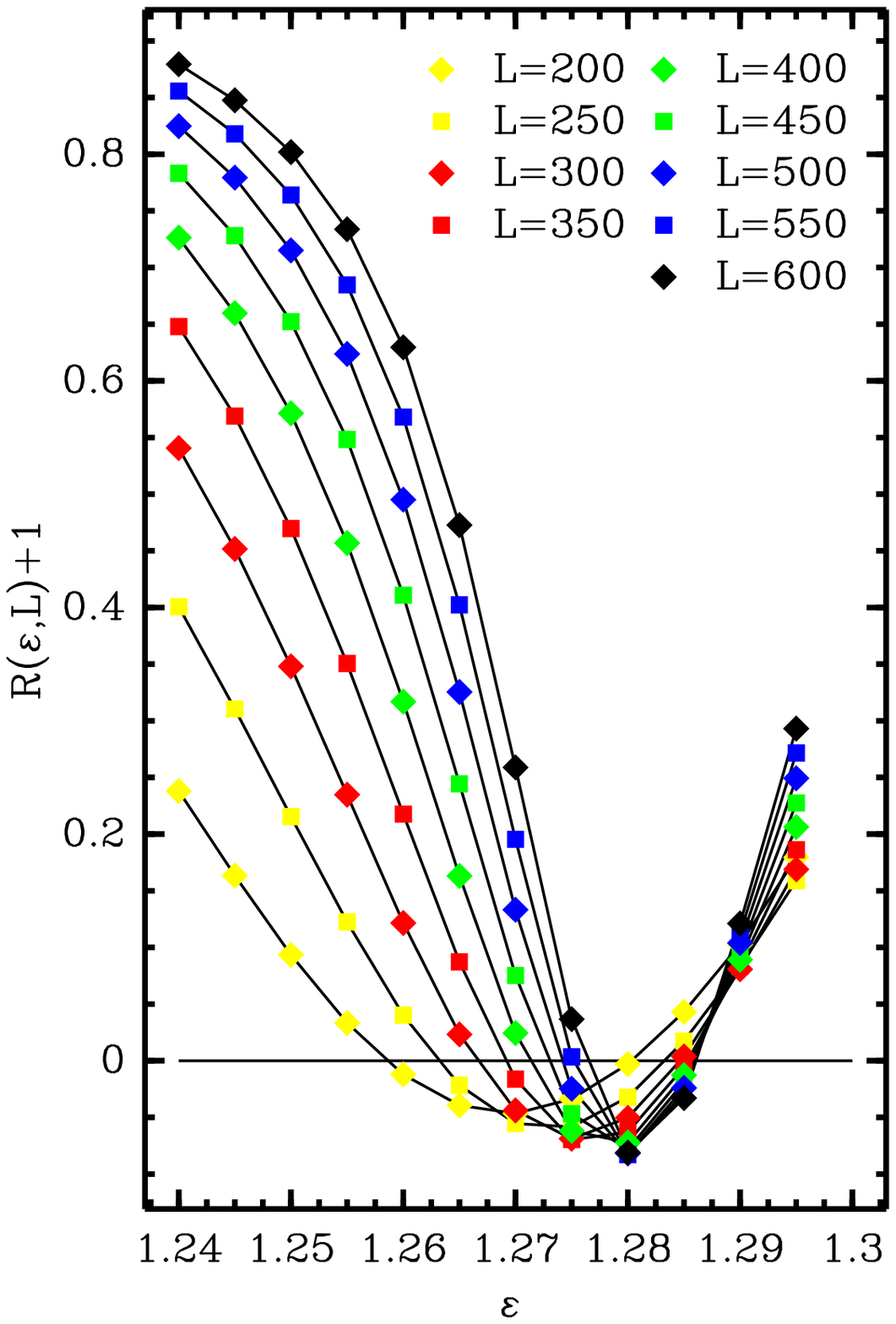}
 \label{fig:pcpA}}
 \subfigure[]{
   \includegraphics[width=\halfcolumnwidth,height=\brcolumnwidth]{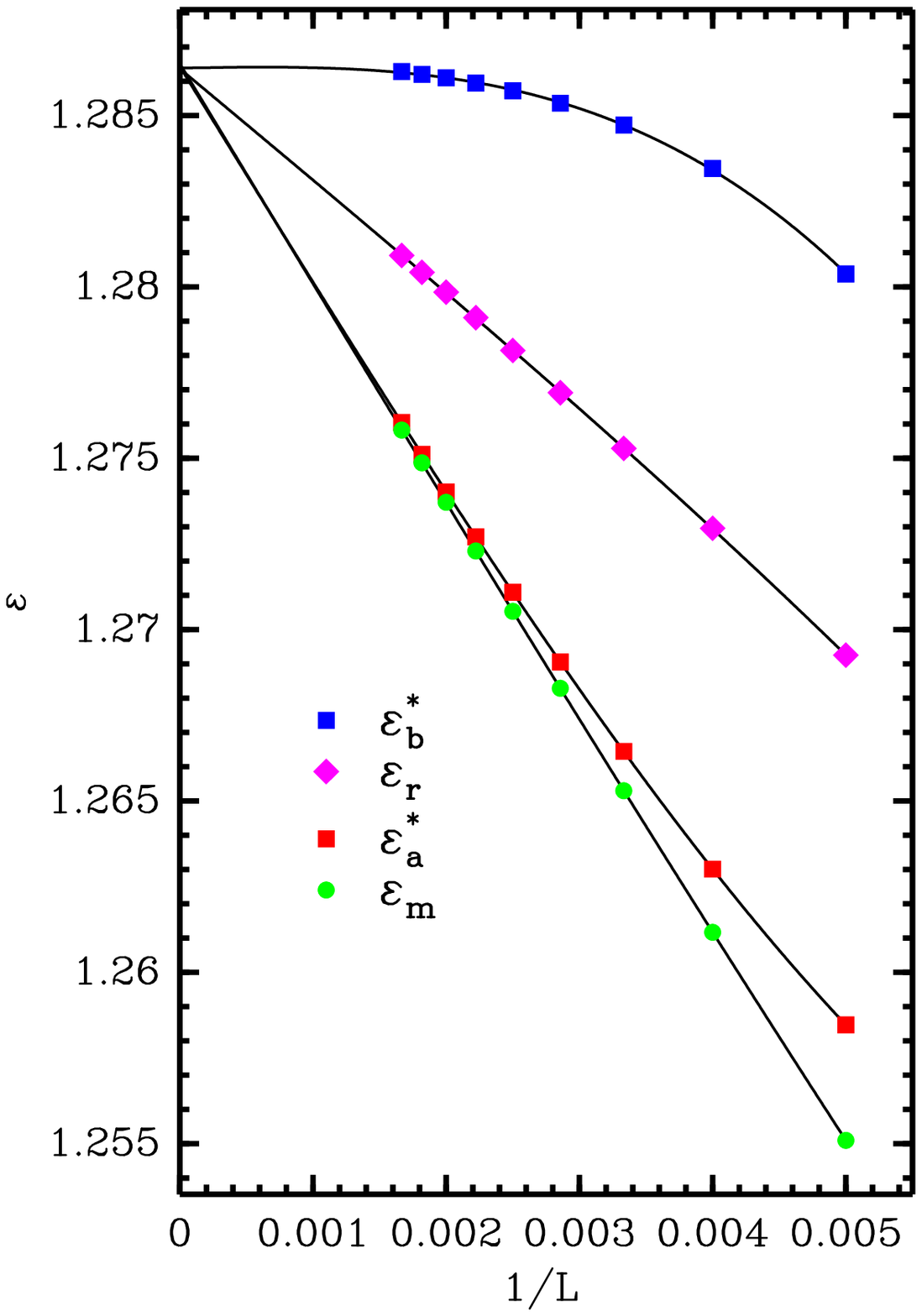}
 \label{fig:pcpB}}
  \caption{(Color online) \subref{fig:pcpA} Logarithmic mass gap ratio 
    as a function of $\varepsilon$ for various system sizes and 
    \subref{fig:pcpB} finite-size extrapolations of the critical point
    using both sequences of pseudo-critical points, as well as 
    the two definitions of the gap minimum. Here $\varepsilon_m$ is 
    the position of the mass gap minimum, $\varepsilon_r$ the position
    of the mass-gap ratio minimum, and $\varepsilon^*_a$ and 
    $\varepsilon^*_b$ are the upper and lower sequences of
    pseudo-critical points, respectively.
    The lines are guides to the eye.}
  \label{fig:pcp}
\end{figure}

In order to proceed, we next need to calculate the critical value of
the coupling in the thermodynamic limit, $\varepsilon_{c_1}$.
The most efficient and accurate way of doing this is to carry out
scaling using the {\em logarithmic} mass gap ratio,\cite{Hamer_1980}
defined as
\begin{equation}
R\left(\varepsilon, L\right) = \frac{ \ln F\left(\varepsilon, L+2\right) - \ln F\left(\varepsilon, L\right) }
{\ln (L+2) - \ln L} \, .
\label{equ:logratio}
\end{equation}
This quantity can be used to define a sequence of pseudo-critical points
for different system sizes using the criterion 
$R\left(\varepsilon^*,L \right) + 1 = 0$.
In Fig.~\ref{fig:pcpA} we show the behavior of $R\left(\varepsilon^*,L \right) + 1$
as a function of $\varepsilon$ for various system sizes.
The curves of the scaled ratio cross the line at two points, 
defining two sets of pseudo-critical points, which we designate as
$\varepsilon_a^*(L)$ and $\varepsilon_b^*(L)$ for the lower and upper
crossings, respectively.
The finite-size scaling of both series of pseudo-critical points 
is depicted in Fig.~\ref{fig:pcpB}. 
All curves are fit 
with third-order polynomials in $1/L$. 
In the thermodynamic limit, $\varepsilon^*_a$ and $\varepsilon^*_b$
converge to the same point to within the accuracy of the extrapolation,
confirming that the transition is second order.
The finite-size scaling of the position of the minimum 
in the mass gap provides an alternate way of determining $\varepsilon_{c_1}$.
This can either be done using the mass gap, Eq.~(\ref{equ:masgap}), directly,
which we designate as $\varepsilon_m(L)$, or using the minimum of the mass-gap ratio, 
Eq.\ (\ref{eq:genmassgapratio}), designated as $\varepsilon_r(L)$.
The extrapolations of positions of the minima, 
$\varepsilon_m$ and $\varepsilon_r$ also converge to the same point, 
providing a confirmation of the consistency 
and stability of the extrapolation procedure.
%
%
%
%
We obtain the location of the critical point at
\begin{equation}
\varepsilon_{c_1} = 1.286(5)
\, .
\label{equ:1CP}
\end{equation}

We can now estimate the correlation-length 
exponent using the finite-size version of the Callan-Symanzik
$\beta$-function~\cite{Hamer_1980,Hamer_1981,Hamer_1981b,Barber_1983}
\begin{equation}
\beta_{\text{cs}}^{-1}\left(\varepsilon, L \right) = \frac{1}{F\left(\varepsilon, L\right)}
\frac{\partial F\left(\varepsilon, L\right)}{\partial \varepsilon} \, ,
\label{eq:betafunction}
\end{equation}
which has critical behavior 
\begin{equation}
\beta_{\text{cs}} \left( \varepsilon_{c_1}, L \right) \sim L^{-\frac{1}{\nu}}
\,.
\label{eq:betapropr}
\end{equation}
To calculate the exponent $\nu$, we proceed as follows: 
Given a sequence of pseudo-critical points, $\varepsilon^*(L)$, we
extrapolate the ratio of the $\beta$-functions for different system sizes
\begin{equation}
\frac{ \beta_{\text{cs}} \left(\varepsilon^*, L + \ell \right) }{ \beta_{\text{cs}} \left(\varepsilon^*, L \right) }
\sim \left( \frac{L+\ell}{L} \right)^{-\frac{1}{\nu}}
\nonumber 
\end{equation}
to the thermodynamic limit. 
Here it is important to choose $\varepsilon^*(L)$ carefully: 
extrapolating using a series of pseudo-critical points
that is close to the gap minimum can yield unreliable
results because the derivative of the mass gap remains zero
or close to zero.
Therefore, we utilize the ratio from the second series 
of pseudo-critical points $\varepsilon^*_L = \varepsilon_b^*(L)$ rather
than from the first $\varepsilon_a^*(L)$ [see Fig.~\ref{fig:pcpB}].
%
%
\begin{figure}[ht]
  \centering
  \subfigure[]{
    \includegraphics[width=\halfcolumnwidth,height=\brcolumnwidth]{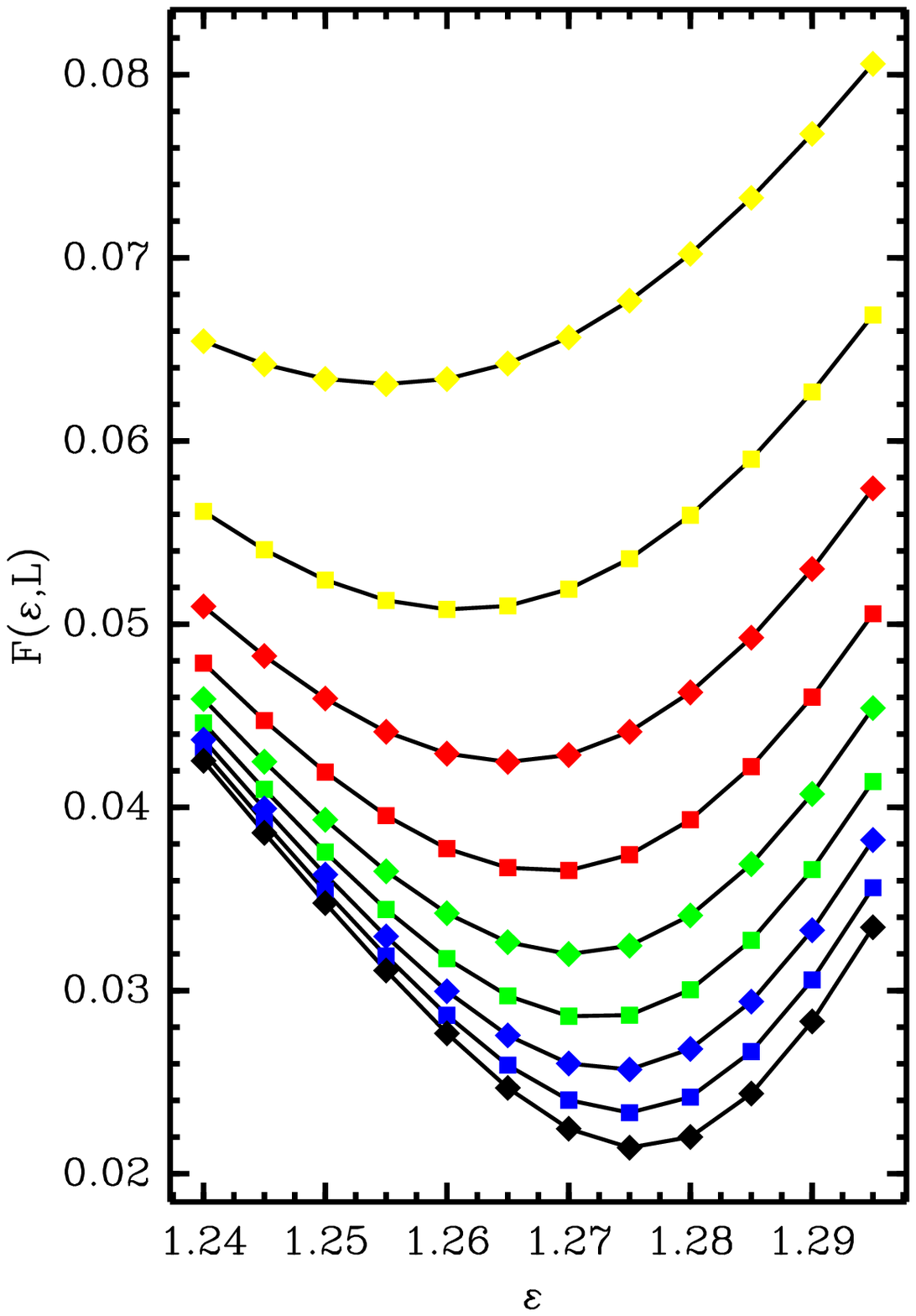}
  \label{fig:massgapA}}
  \subfigure[]{
    \includegraphics[width=\halfcolumnwidth,height=\brcolumnwidth]{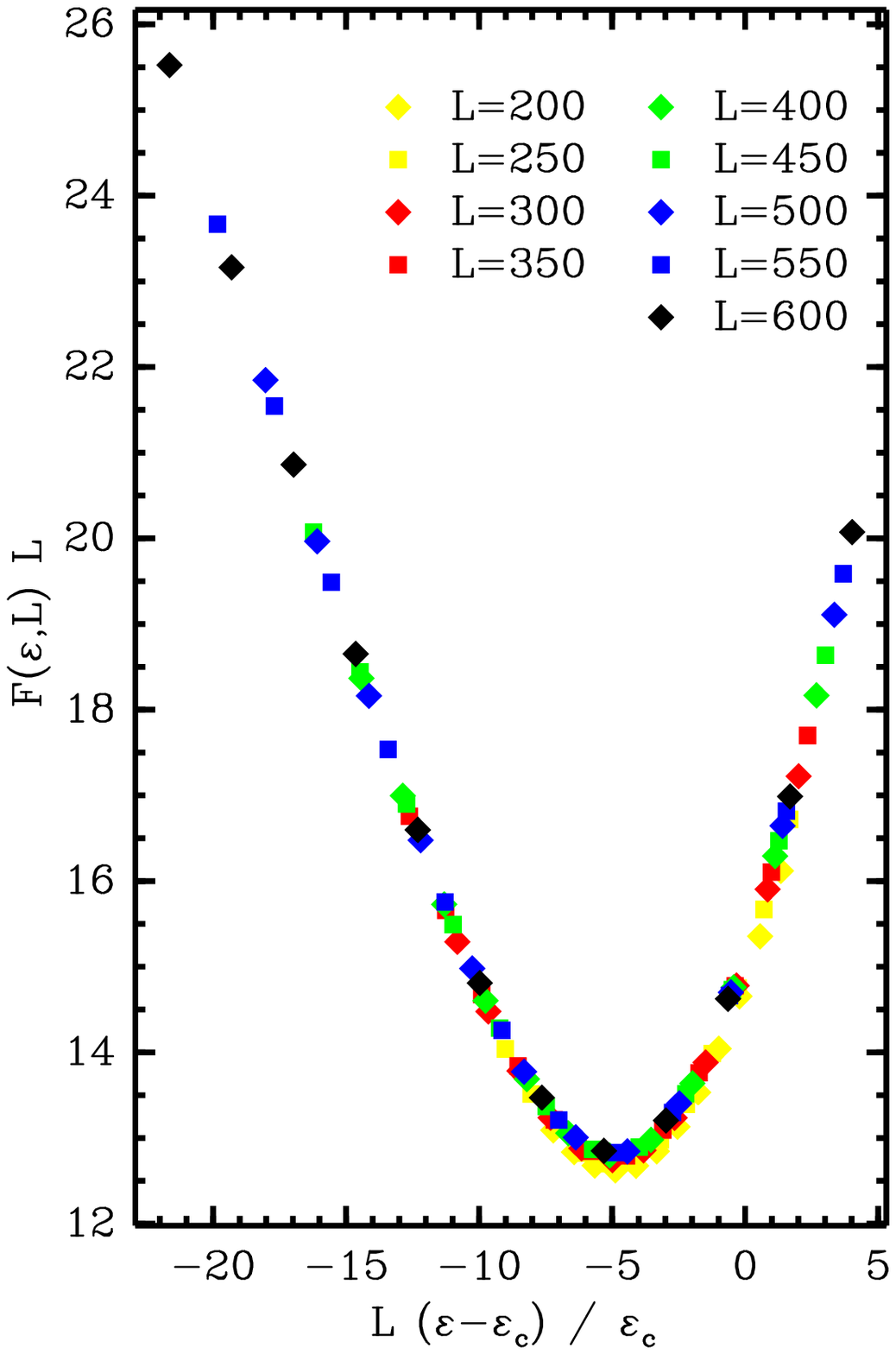}
  \label{fig:massgapB}}
  \caption{ (Color online) Scaling of the mass gap around the first critical point:
    \subref{fig:massgapA} unscaled data.  The lines are guides to the eye. 
    \subref{fig:massgapB} Rescaled data
    $F \left(\varepsilon, L \right) L$ are plotted as a function
    of the rescaled coupling $L \left(\varepsilon-\varepsilon_{c_1}\right)/\varepsilon_{c_1}$.}
  \label{fig:massgap}
\end{figure}
%
%
If $|1/\nu| < 1$ and $L >> \ell$, then
\begin{equation}
\frac{L}{\ell} \left[ \frac{ \beta_{\text{cs}} \left(\varepsilon^*, L+\ell \right) }
{ \beta_{\text{cs}} \left(\varepsilon^*, L\right) } - 1 \right]
\sim - \frac{1}{\nu} 
\,.
\nonumber
\end{equation}
From the numerical extrapolation, we obtain
\begin{equation}
\frac{1}{\nu} = 0.996(5)
\, .
\label{equ:nuexponent}
\end{equation}
A plot of unscaled mass-gap data and its collapse using this scaling
exponent is shown in Fig.~\ref{fig:massgap}. 
\subsection{Thermodynamic exponents $\beta$, $\alpha$, $\gamma$}
\begin{figure}[htb]
  \centering
  \subfigure[]{
    \includegraphics[width=\halfcolumnwidth,height=\brcolumnwidth]{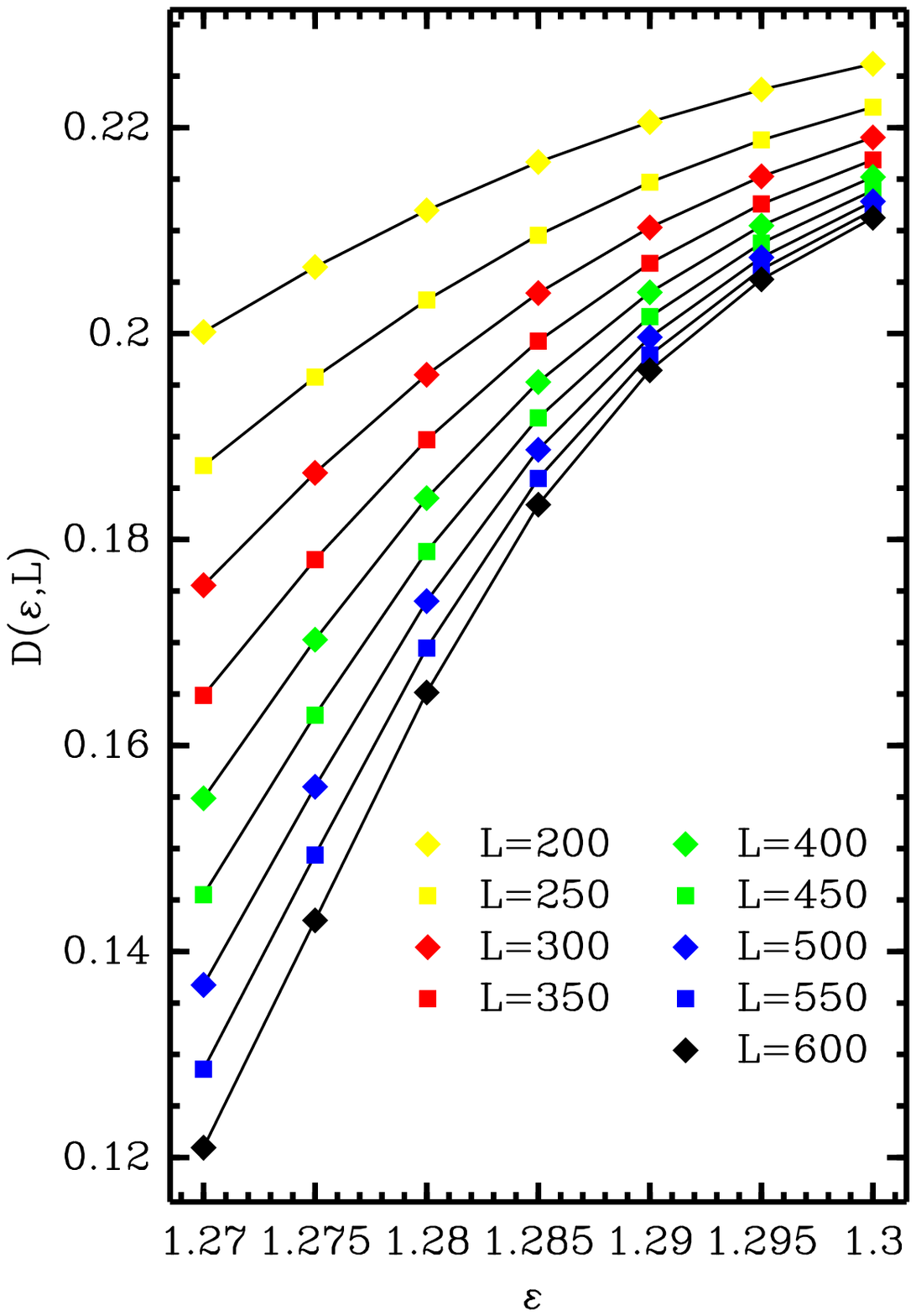}
  \label{fig:BondA}}
  \subfigure[]{
    \includegraphics[width=\halfcolumnwidth,height=\brcolumnwidth]{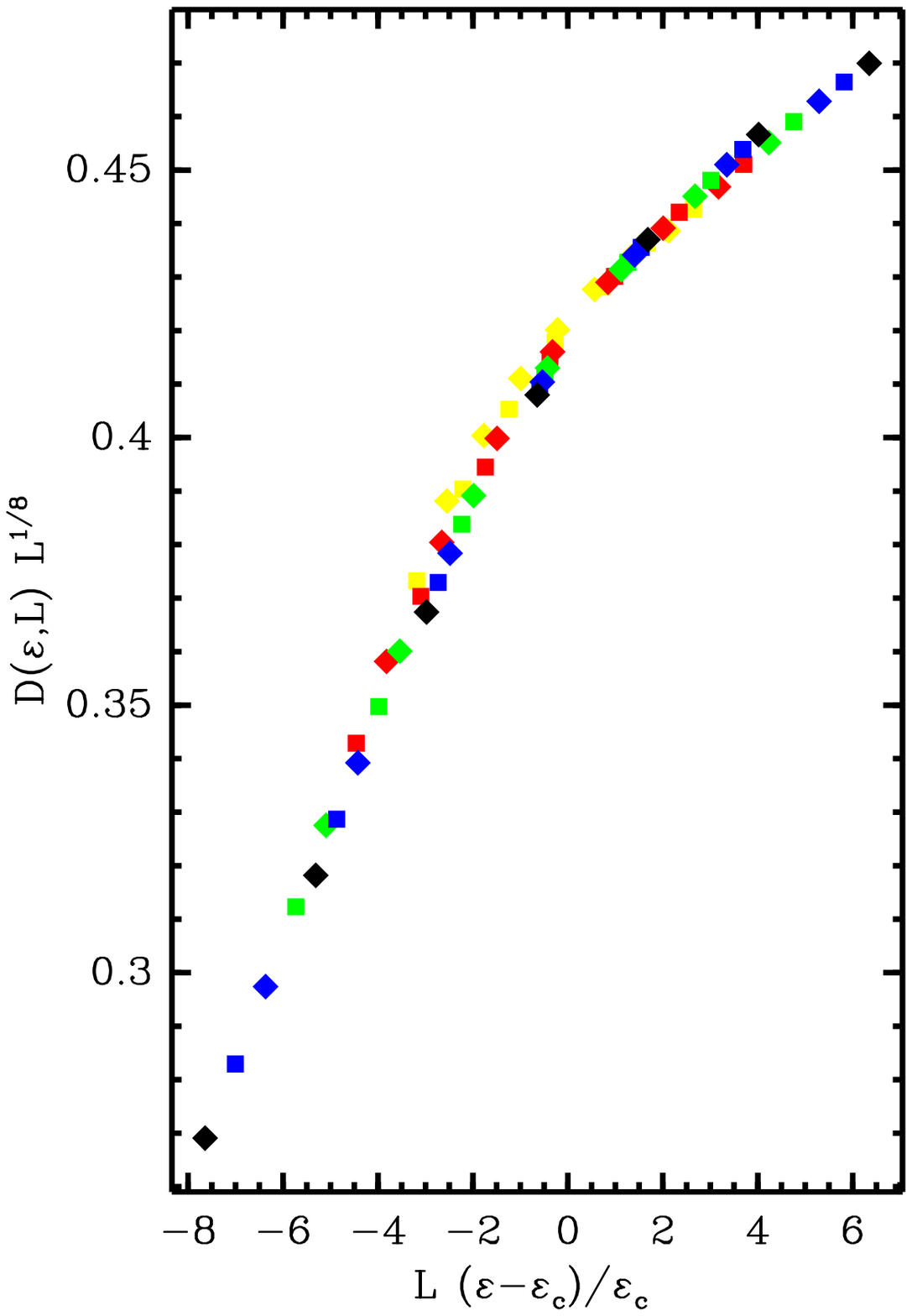}
  \label{fig:BondB}}
  \caption{(Color online) Bond order parameter as a function 
    of the coupling $\varepsilon$ around the transition point:
    \subref{fig:BondA} data for different system
    sizes near the first transition and 
    \subref{fig:BondB} data rescaled as
    $D \left( \varepsilon, L \right) L^{1/8}$  plotted as a function of the rescaled coupling
    $L \left( \varepsilon - \varepsilon_{c_1} \right) /  \varepsilon_{c_1}$.}
  \label{fig:Bond}
\end{figure}

\begin{figure}[htb]
  \centering
  \includegraphics[width=\mycolumnwidth,height=\brcolumnwidth]{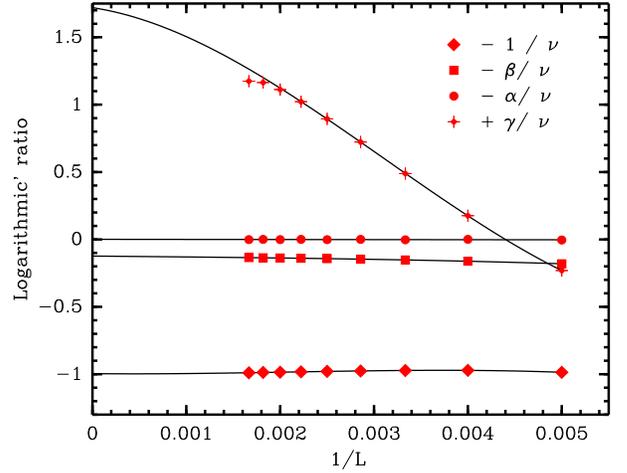}
  \caption{(Color online)
    Finite-size behavior of the exponents $\nu$, $\beta$, $\alpha$, and $\gamma$.
    The fit is to a third-degree polynomial in $1/L$. 
    Note that the points for the two smallest $1/L$ are not included in fitting $\gamma$.}
  \label{fig:exponents}
\end{figure}

The bond order parameter characterizes the bond-order-wave (BOW) phase.
Fabrizio, Gogolin, and Nersesyan have
argued that the bond order parameter is
the right quantity to characterize the Ising transition 
in the IHM.~\cite{Fabrizio_1999,Fabrizio_2000}
The order parameter, 
expressed in the spin-one language, is given by
\begin{equation}
D \left( \varepsilon, L \right) = 
\frac{1}{L-1}
\sum_{i=1}^L  \left(-1\right)^i 
\left[
\left \langle 
\left(\Sop^{+}_i\Sop^{-}_{i+1} + \Sop^{-}_i \Sop^{+}_{i+1}\right) 
\Sop^{z}_{i+1} 
\right \rangle 
\right.
\nonumber
\end{equation}
\begin{equation}
\left. 
- \left \langle  
\Sop^{z}_{i} 
\left( \Sop^{+}_i \Sop^{-}_{i+1} + \Sop^{-}_{i}\Sop^{+}_{i+1}\right)
\right \rangle 
\right] 
\,.
\label{equ:bop}
\end{equation}
DMRG results for the bond order parameter as a function 
of the coupling $\varepsilon$ near the first transition point are depicted
for various system sizes in Fig.~\ref{fig:BondA}.
%
%
%
\newline
We can use the bond order parameter to determine the associated
critical exponent $\beta$, i.e.,
\begin{equation}
D \left(\varepsilon \sim \varepsilon_{c_1}, L \right) 
\sim 
L^{-\frac{\beta}{\nu}} 
\, .
\label{equ:boplimit}
\end{equation}
Using the logarithmic derivative
\begin{equation}
\frac{ \ln {D \left(\varepsilon^*, L + \ell \right) } - \ln{ D \left(\varepsilon^*, L \right)}}
{\ln \left(L+\ell\right) - \ln{L}}
\sim 
- \frac{\beta}{\nu} 
\, ,
\end{equation}
we obtain
\begin{equation}
\frac{\beta}{\nu} = 0.124(5)
\, .
\label{equ:betaexp}
\end{equation}
The excellent data collapse of the rescaled data, as can be seen in
Fig.~\ref{fig:BondB},
confirms that the transition point belongs 
to the 2D Ising universality class.
Results for the finite-size scaling of the exponent $\beta$ are plotted in
Fig.~\ref{fig:exponents}.
%
%
%
%

%
Since, 
in a quantum phase transition,
the coupling plays the same role
as temperature in a thermal phase transition,
we can define a corresponding ``specific heat'' \cite{Kogut_1979,Hamer_1983}
\begin{equation}
c_v \left( \varepsilon, L \right) = - \frac{\varepsilon}{L} 
\frac{\partial^2 E_0\left(\varepsilon, L\right)}{\partial \varepsilon^ 2}
\,.
\nonumber
\end{equation}
Note that this quantity does not correspond to the real specific heat.
Nevertheless, 
due to  the scaling relations and its interplay with the other quantities, 
it has to diverge with the exponent $\alpha$.
The physical specific heat exponent
is related to our $\alpha$ by the Gr{\"u}neisen parameter.~\cite{Zhu_2003}
The specific heat usually contains a regular term 
that is typically larger in amplitude than the singular one.
Therefore, 
instead of using the logarithmic derivative to estimate the exponent $\alpha / \nu$,
we instead use the ratio
\begin{equation}
\frac{L}{2} 
\frac{c_v\left(\varepsilon, L+2\right) - c_v\left(\varepsilon, L\right)}{c_v\left(\varepsilon, L\right)} 
\sim \frac{\alpha}{\nu}
\, .
\end{equation}
\newline
To overcome possible problems in determining this exponent, 
we use the Hellman-Feynman~\cite{Feynman_1939}
theorem to exploit the accuracy of the DMRG in calculating local quantities
\begin{equation}
\frac{\partial E_0\left( \varepsilon, L\right)}{\partial \varepsilon } = 
- \frac12
\sum_i^L \left \langle \left(\Sop_i^{z} \right)^{2} \right \rangle 
\, .
\end{equation}
This trick reduces the computational cost
to that of calculating the first derivative of the cubic spline, 
which interpolates the data points.~\cite{Press_1999}
The result is the following:
\begin{equation}
\frac{\alpha}{\nu} = 0.00(1)
\, .
\label{equ:alphaexp}
\end{equation}
\newline
The finite-size behavior of the various exponents 
is plotted in Fig.~\ref{fig:exponents}.
The scaling relation $\alpha = 2 \left(1 - \nu\right)$ is fulfilled
by Eqs.~(\ref{equ:nuexponent}) and (\ref{equ:alphaexp}).~\cite{Barber_1983}
Finally, 
we determine the exponent $\gamma$ associated with the relevant susceptibility.
The susceptibility corresponding to the bond order parameter is
\begin{equation}
\chi_{D} \left(\varepsilon, L \right) = 
- \frac{1}{L} \left. 
\frac{\partial D \left(\varepsilon, L\right)}{\partial h_{D}} \right|_{h_{D}=0} .
\end{equation}
In order to calculate this quantity, 
we turn once more to the Hellman-Feynman
theorem and to linear response theory.
We perturb the Hamiltonian 
with a small field $h_{D}$ conjugate to the order parameter $D$. 
The field has to be small enough to reveal a linear
regime in the changes, but not smaller than the actual DMRG resolution;
we use $2 \delta h_D = 10^{-4} t $.
We have measured the order parameter for four points around $h_D=0$
in order to compute its first derivative at $h_D=0$.
%
%
\begin{figure}[htb]
  \centering
  \subfigure[]{
    \includegraphics[width=\halfcolumnwidth,height=\brcolumnwidth]{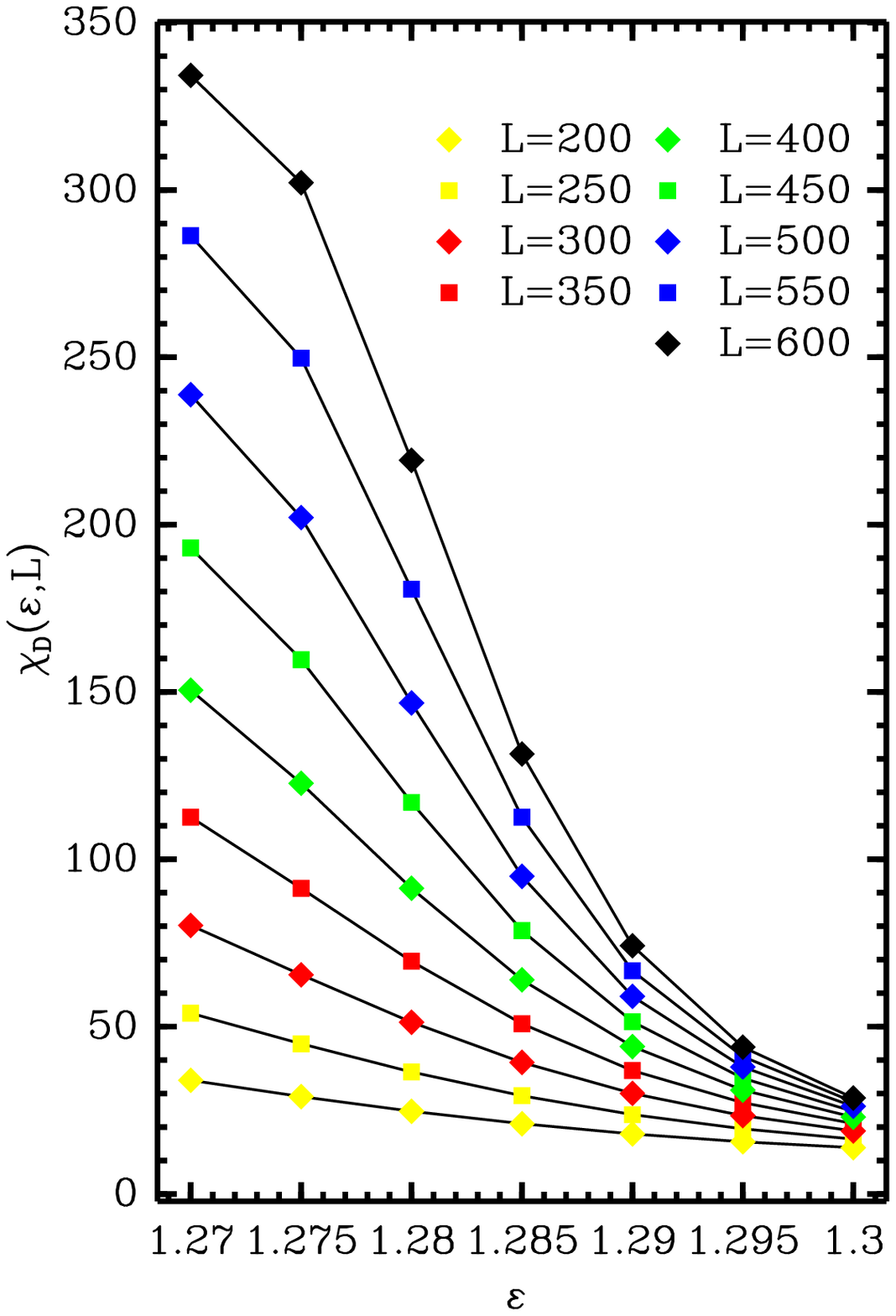}
  \label{fig:rbopsA}}
  \subfigure[]{
    \includegraphics[width=\halfcolumnwidth,height=\brcolumnwidth]{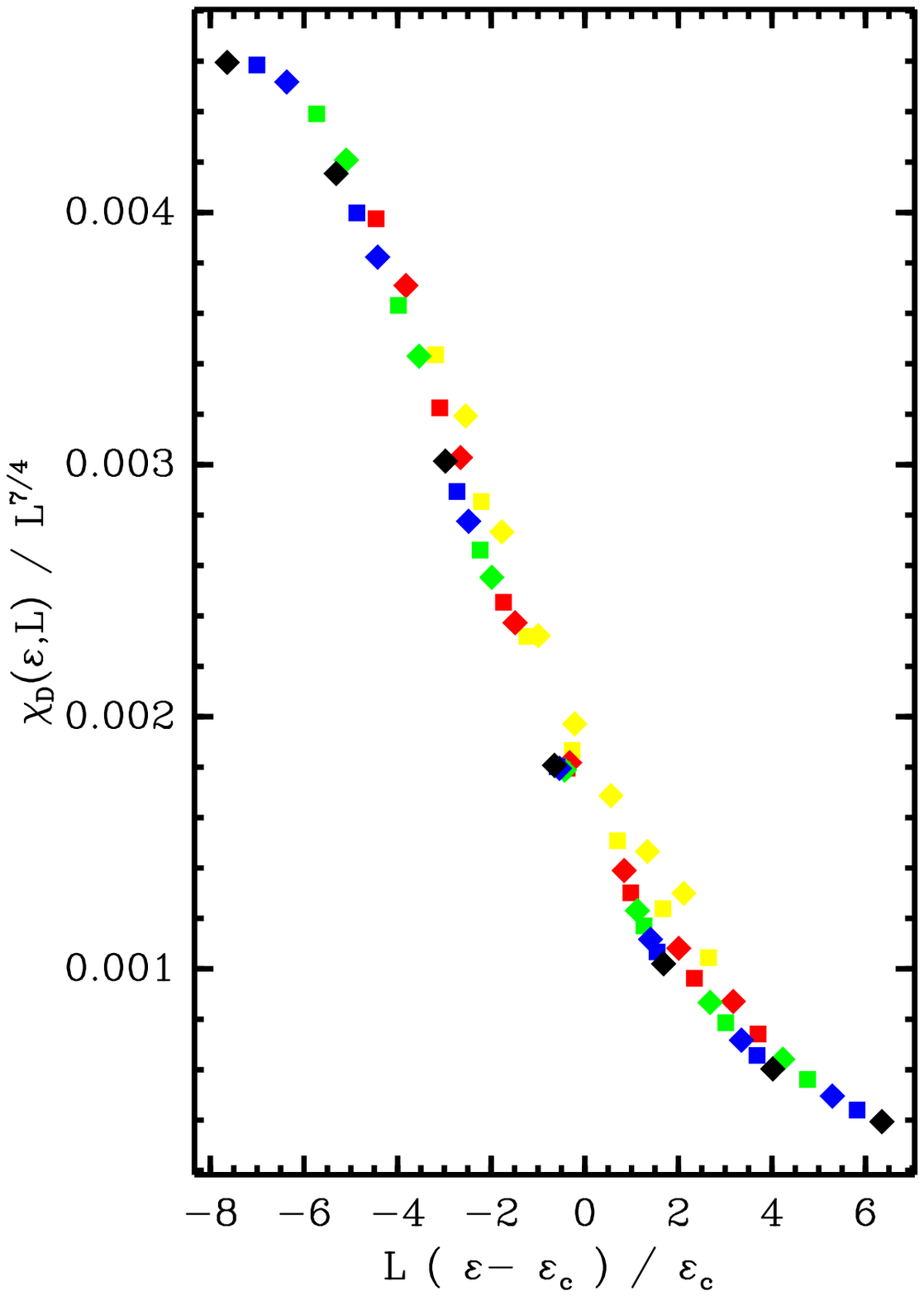}
  \label{fig:rbopsB}}
  \caption{(Color online)
    \subref{fig:rbopsA} Bond order parameter susceptibility
    as function of the coupling $\varepsilon$ for different system sizes.
    \subref{fig:rbopsB} The collapsed curves scaled using the exponent $\gamma=7/4$.}
  \label{fig:rbops}
\end{figure}
%
%

%
Once we have evaluated 
the static susceptibility for different system sizes, 
we proceed in the same way as for the previous exponents. 
The scaling relation is
\begin{equation}
\chi_D \left(\varepsilon_{c_1}, L\right) 
\sim L^{\gamma/ \nu}
\, .
\end{equation}
Thus, from
\begin{equation}
\frac{ \ln {\chi_D \left( \varepsilon^*, L+\ell \right)}- \ln{ \chi_D \left(\varepsilon^*, L \right) }}
{\ln \left( L + \ell \right) - \ln{L}}
\sim \frac{\gamma}{\nu}
\label{eq:gammascal} 
\end{equation}
we obtain the last thermodynamic exponent,
as plotted in Fig.~\ref{fig:exponents}, 
with the value
\begin{equation}
\frac{\gamma}{\nu} = 1.72(5)
\, .
\label{equ:gammaexp}
\end{equation}
As shown in the figure, 
the last points for the largest system sizes
have been excluded in calculating the exponent. 
The reason is that 
the calculation of the susceptibility becomes uncontrolled 
for very big system sizes.
In order to compensate the occurrence of nonlinear behavior 
in the response for larger system sizes, 
we would have to use a very small perturbation field. 
However, the effect of such a small field can be difficult 
to distinguish from the numerical noise.
In addition, we have to carry out two cubic-spline interpolations:
one to determine the derivative of the bond order parameter 
as function of the perturbation field and one to fit its susceptibility.
For these reasons we neglect the points at the two largest system sizes.
We see that the second scaling relation $\gamma = 2 \left(\nu - \beta\right)$
is fulfilled to within our estimated error.~\cite{Barber_1983}
Other quantities, 
such as the electric polarization and the electric susceptibility, 
scale with the same exponents
as the bond order parameter 
and the bond-order susceptibility, respectively.~\cite{Tincani_2008}
In addition, 
we have calculated the value of the central charge governing the
underlying conformal field theory numerically in two different ways.
In the first method, we use that 
the scaling of the low-lying energy levels with
system size is uniquely determined by the conformal tower.\cite{Boschi_2004}
This scaling can be used to determine the central charge.~\cite{DiFrancesco_1999}
The value obtained, $c=0.50(4)$, 
is consistent with that expected for the 2D Ising model.
In the second method, we determine the central charge from the
entropy profile, which has a known form dependent only on the central
charge.\cite{Calabrese_2004}
We obtain the same value ($c\approx 0.5$) to within the numerical
accuracy at $\varepsilon_{c_1}$.
%


%
\section{\label{sec:sdimi} SDI to MI transition}
In this section,
we present numerical
results on the second transition
where the system passes from
the spontaneously dimerized phase
to the Mott insulator phase
with increasing $\varepsilon$.
We will show both how the spin gap closes when approaching the
critical point $\varepsilon_{c_2}$ from below and how the bond
susceptibility diverges when approaching $\varepsilon_{c_2}$ from
above.
Our results confirm the KT scenario with an essential singularity at
$\varepsilon_{c_2}$ and a critical phase for 
$\varepsilon > \varepsilon_{c_2}$.
In order to do this, a more careful treatment 
than at the 2D Ising transition point is required.
\subsection{Correlation length and mass gap}
\begin{figure}[htbp]
  \centering
  \subfigure[]{
    \includegraphics[width=\halfcolumnwidth,height=\brcolumnwidth]{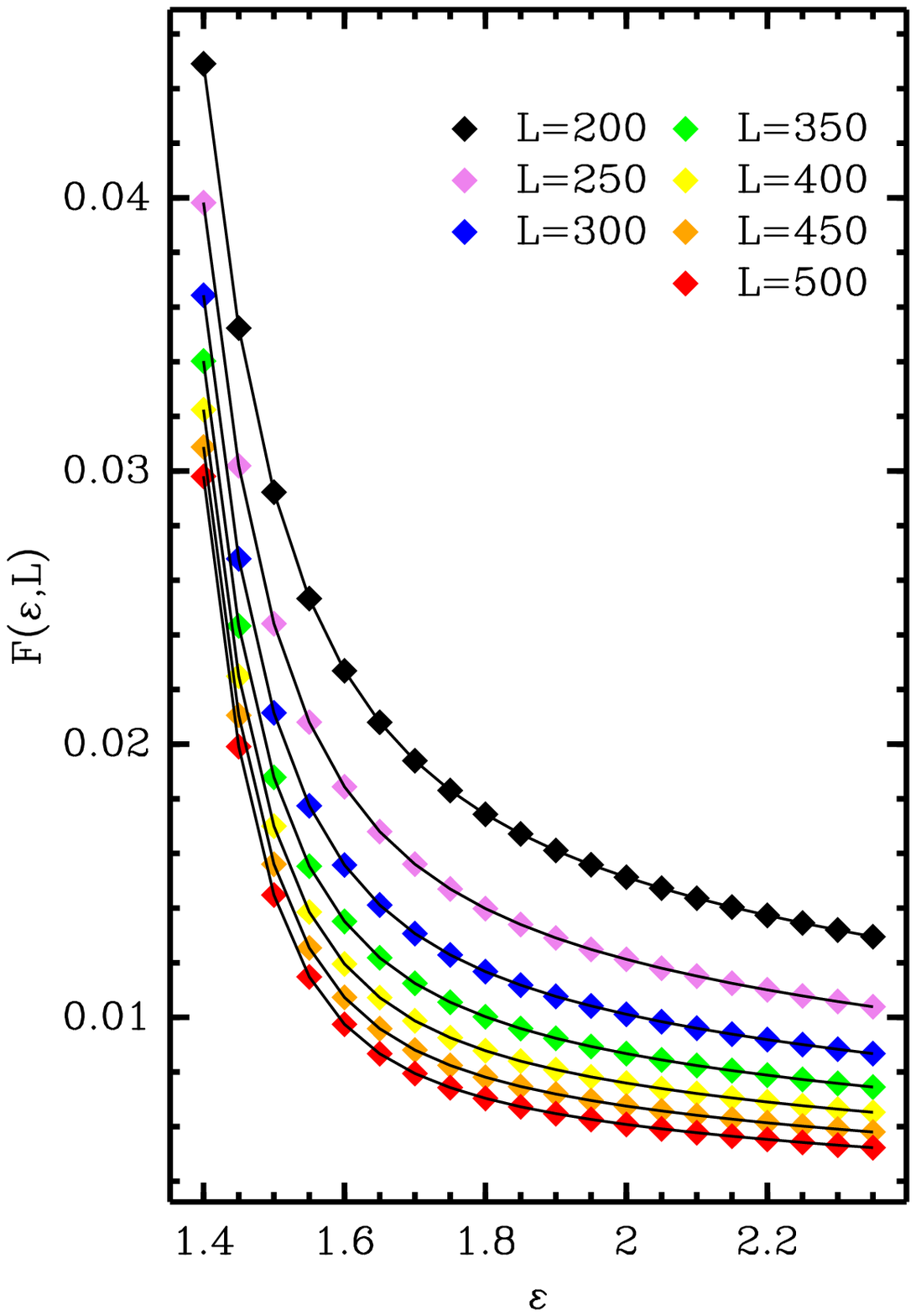}
  \label{fig:mgr-tk-tlA}}
  \subfigure[]{
    \includegraphics[width=\halfcolumnwidth,height=\brcolumnwidth]{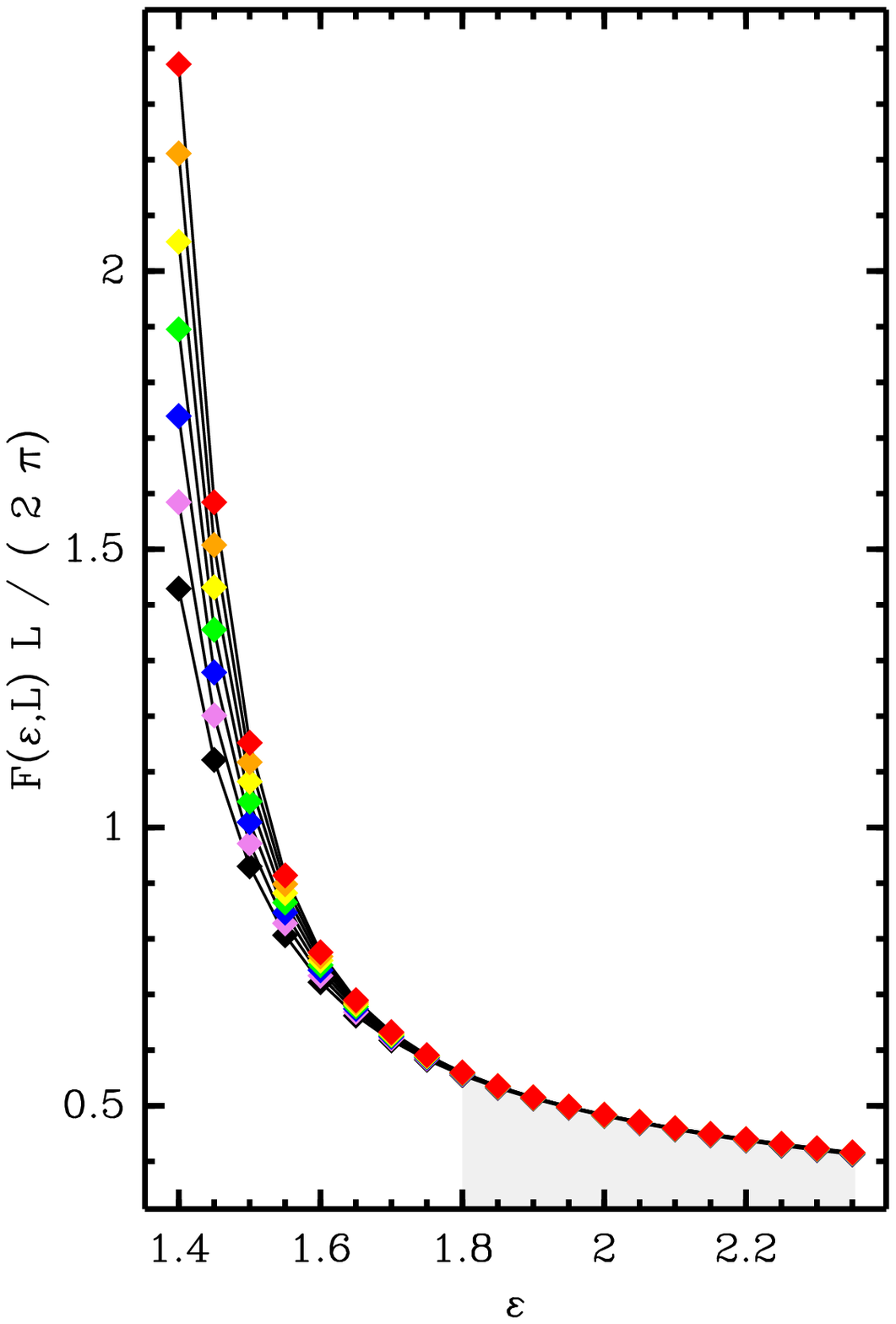}
  \label{fig:mgr-tk-tlB}}
  \caption{(Color online)
    \subref{fig:mgr-tk-tlA} Mass gap and \subref{fig:mgr-tk-tlB} mass gap times $L$ 
    relative to the second critical point as function of the coupling $\varepsilon$ 
    and for different system sizes.}
  \label{fig:mgr-tk-tl}
\end{figure}

\begin{figure}
  \centering
  \subfigure[]{
    \includegraphics[width=\mycolumnwidth]{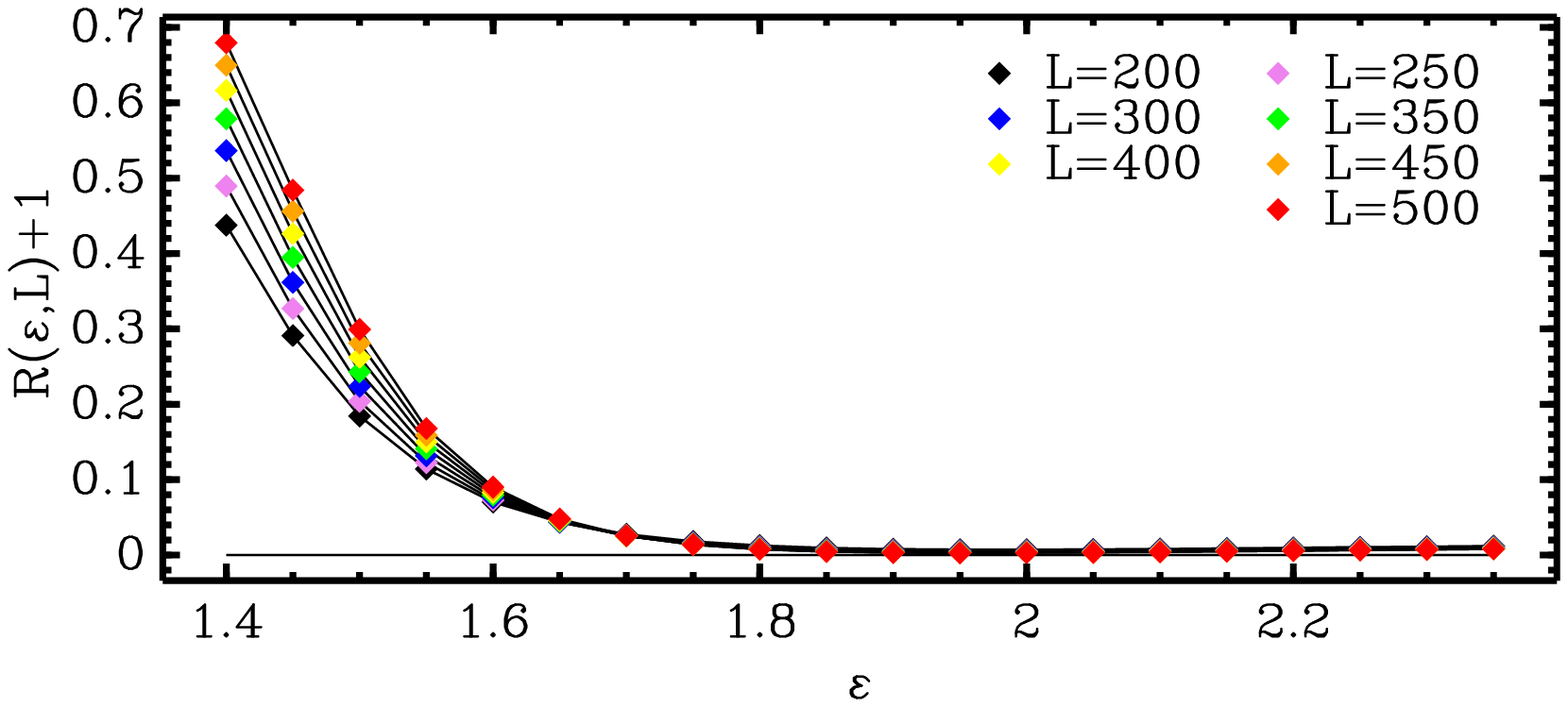}
  \label{fig:mgr-ratio-tkA}}
  \subfigure[]{
    \includegraphics[width=\mycolumnwidth]{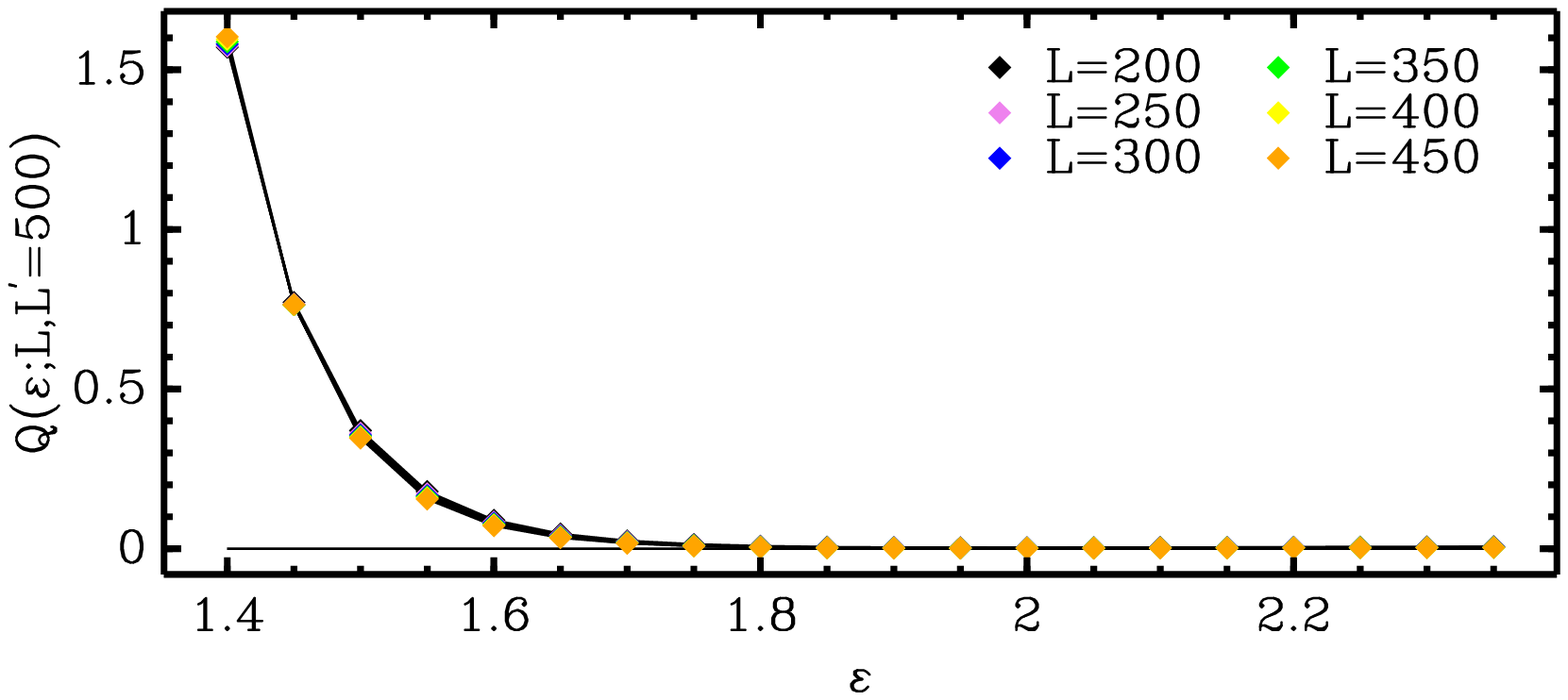}
  \label{fig:mgr-ratio-tkB}}
  \caption{(Color online)
    \subref{fig:mgr-ratio-tkA} The logarithmic mass gap ratio plus unity.
    \subref{fig:mgr-ratio-tkB} The scaled difference of mass gaps Q for $L^\prime = 500$ 
    and various values of $L$.}
  \label{fig:mgr-ratio-tk}
\end{figure}

For a KT transition,
the correlation length diverges exponentially 
as the transition point is approached from the gapped phase
and remains infinite in the critical region that follows.~\cite{Kosterlitz_1974}
Since the mass gap is related to the inverse of the correlation length,
the mass gap has to close exponentially as the transition is approached
and is zero in the critical region.
However, 
for finite-size systems, 
the correlation length $\xi$ is limited by the system size $L$. 
Very large system sizes or the inclusion of higher
order corrections are required to reveal the exponential divergence,
which is restricted to a narrow region close to the KT transition.
In order to locate the position 
of the second transition point $\varepsilon_{c_2}$,
we analyze the scaling of the mass gap, depicted in Fig.~\ref{fig:mgr-tk-tlA}.
The finite-size scaling analysis
made for the first transition
cannot be used here
because sufficiently large systems 
to study the logarithmic scaling cannot be reached.
Instead, 
we prefer to use a different approach
based on conformal field theory (CFT).
Within the Mott insulator phase, where the spin sector is gapless,
the system is critical and can be described by a CFT.
Furthermore,
the characteristic excitation gaps scale with system size $L$ as 
\begin{equation}
E_i\left( L \right) - E_0 \left( L \right)
= \frac{2 \pi x_i v }{L} \, ,
\label{eq:cftratio}
\end{equation}
where $x_i$ is the corresponding scaling index and $v$ 
is the ``excitation'' velocity.
Since
this expression is valid only in the critical region
corresponding to the Mott insulator,
the extent to which it is fulfilled can be used
to locate the transition point.
In a plot of the mass gap times the system size $L$, Fig.~\ref{fig:mgr-tk-tlB}, 
all curves merge into a single one exactly at a critical point
$\varepsilon_{c_2}$, as expected from Eq.~(\ref{eq:cftratio}).
%
%
%
%
Therefore, 
the system is in a critical regime above
a critical coupling 
\begin{equation}
\varepsilon_{c_2} = 1.8(1)
\label{equ:2CP}
\, .
\end{equation}
The point at which the curves merge is clearly separated,
see Eq.~(\ref{equ:1CP}),
from the first critical point that we found at $\varepsilon_{c_1}$.
An analysis of the mass gap ratio,
see Eq.~(\ref{equ:logratio}), is also useful.
In contrast to what happens at the first transition, $\varepsilon_{c_1}$,
the curves do not cross the line corresponding to a ratio of unity
due to the logarithmic corrections.~\cite{Hamer_1981b}
Nevertheless, 
the curves remain very close to zero everywhere 
in the critical region above $\varepsilon_{c_2}$,
as can be seen in Fig.~\ref{fig:mgr-ratio-tkA}.
In the region preceding $\varepsilon_{c_2}$, 
the value of the mass-gap ratio increases with the system size,
as expected for a gapped system.
The overall behavior of the mass-gap ratio curves 
further confirms that there is a second transition point 
at $\varepsilon_{c_2}$ and supports the KT scenario.
In addition,
we define and calculate 
the scaled difference of mass gaps, $Q$,
\begin{equation}
Q \left(\varepsilon; L^\prime,L \right) = 
\frac{L^\prime}{2 \pi} 
\frac{F \left(\varepsilon, L^\prime\right) \cdot L^\prime - F \left(\varepsilon, L\right) 
\cdot L }
{L^\prime - L} 
\,.
\end{equation}
For an arbitrary $L^\prime$,
the first-order finite-size scaling terms cancel out 
and $Q\left( \varepsilon; L^\prime, L\right)$ vanishes in the critical region.
In Fig.~\ref{fig:mgr-ratio-tkB}
we show results for $L^\prime =500$.
We conclude that the second critical point occurs at
$\varepsilon_{c_2} \approx 1.8$
and the gap closes exponentially.
Since the distance between the two critical points
is much bigger, $\varepsilon_{c_2} - \varepsilon_{c_1} \approx 0.5$,
than any deviation due to the logarithmic corrections, 
we conclude that there are two phase transitions.
%
%
%
%

%
We have also calculated the approximate $\beta$-function $\beta_{cs}$,
Eq.~(\ref{eq:betafunction}).
However, 
for this kind of transition, it has no zeros (as expected).~\cite{Hamer_1981b}
Nevertheless, 
we can extrapolate the value of the minima of the $\beta$-function
as a function of the system size
to the thermodynamic limit.
This yields an alternate estimate of $\varepsilon_{c_2}$,
$\varepsilon_{c_2}^\beta = 1.9(1)$.
\subsection{The bond-order and electric susceptibility}
\begin{figure}
  \centering
  \subfigure[]{
    \includegraphics[width=\halfcolumnwidth,height=\brcolumnwidth]{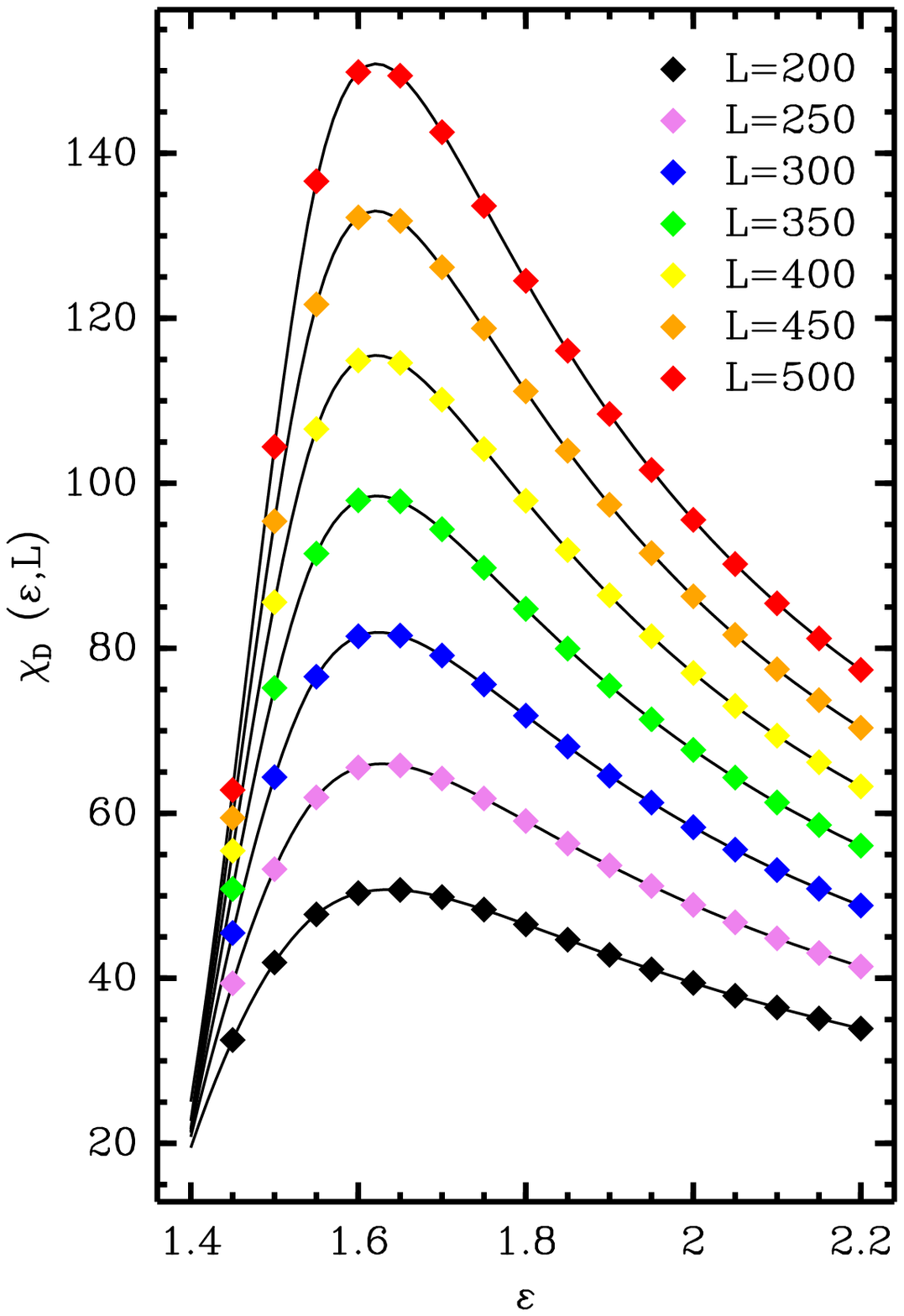}
  \label{fig:bond_susc_tkA}}
  \subfigure[]{
    \includegraphics[width=\halfcolumnwidth,height=\brcolumnwidth]{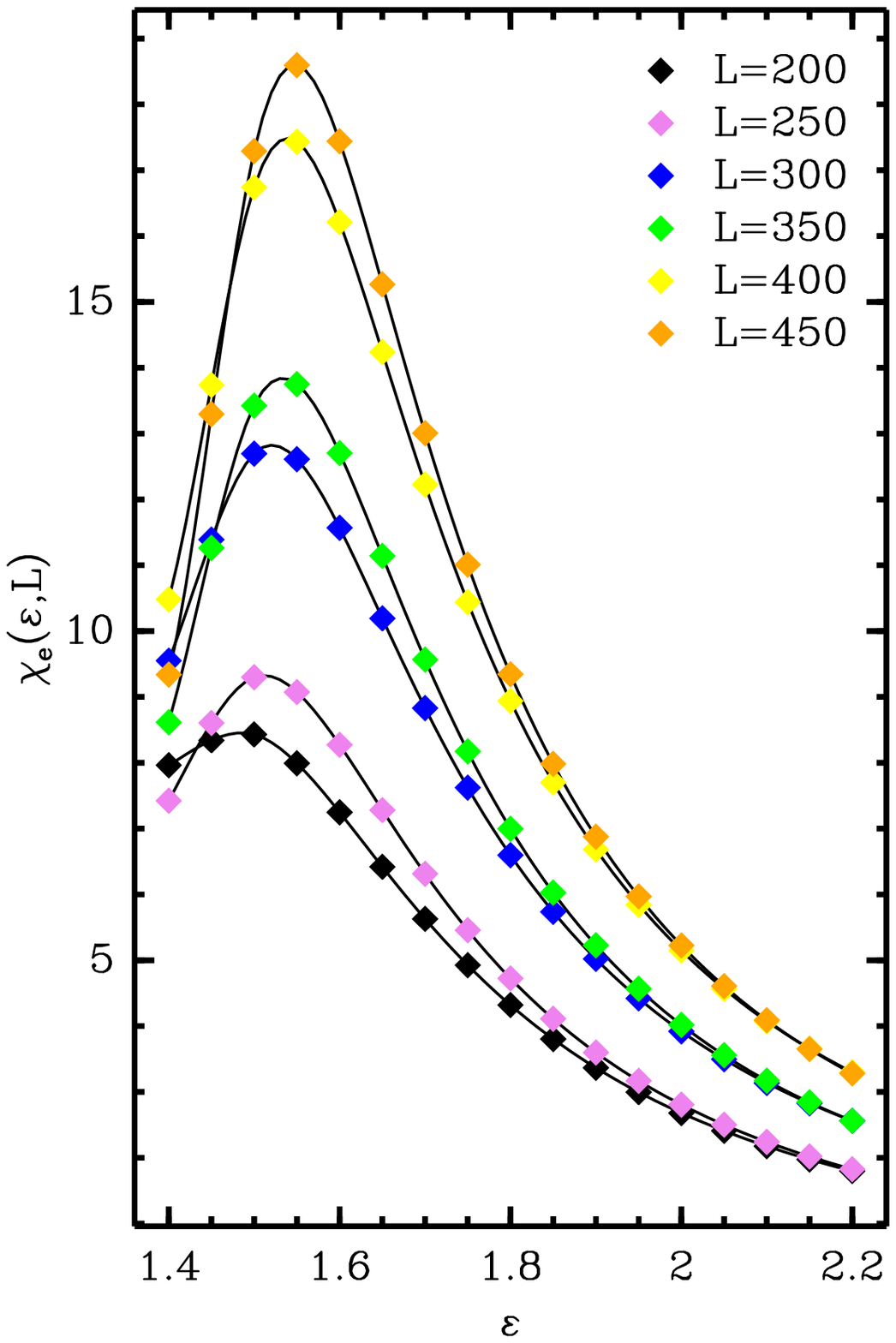}
  \label{fig:bond_susc_tkB}}
  \caption{(Color online)
    \subref{fig:bond_susc_tkA} The bond-order susceptibility and 
    \subref{fig:bond_susc_tkB} the electrical susceptibility 
    for the SDI-MI transition.}
  \label{fig:bond_susc_tk}
\end{figure}

In order to classify the transition as a KT transition, 
we examine the bond-order susceptibility
and the electric susceptibility, see Fig.~\ref{fig:bond_susc_tk}.
%
%
%
%
The behavior of the peak of the bond-order susceptibility 
can be used to estimate the exponent of the susceptibility,
\begin{equation}
\gamma^*_{\text{peak}}\left( L \right) = 
\frac{\ln \chi\left( \varepsilon^*_{\text {peak}}, L +2 \right) - \ln \chi \left( \varepsilon^*_{\text {peak}}, L \right)}
{ \ln \left(L+2 \right) - \ln L} 
\, .
\end{equation}
The position of the peak in the bond-order susceptibility 
converges to the value $\varepsilon_{\rm peak} \approx 1.62$,
and the series of pseudo-exponents, $\gamma^*_{\rm peak}\left( L \right)$, 
converges to $\gamma \approx 1.27$ in the thermodynamic limit.
For comparison, 
we calculate the electric susceptibility, 
shown in Fig.~\ref{fig:bond_susc_tkB}.
The finite-size effects are much stronger for the electric susceptibility
than for the bond-order susceptibility.
In fact, 
we also observe a narrow peak in $\chi_e$ that grows and moves with the system size.
In general, 
we conclude that 
the coincidence of the mass gap closing to zero exponentially
and a diverging susceptibility 
corresponds to the typical scenario of an infinite-order phase transition.
The critical exponent of the susceptibility $\gamma$ 
cannot be determined accurately because of strong finite-size effects 
due to the strong influence of the bond-order wave which scales to zero very slowly,
i.e.,  as $1/\sqrt{L}$.~\cite{Tincani_2008}
Additionally,
we have made a preliminary calculation 
of the central charge from the entropy profile
assuming that it has the form predicted by CFT.~\cite{Calabrese_2004}
In order to determine the transition point, we minimize the $\chi^2$
of the fit to the conformal form and confirm that $c \approx 1$
($c=1$ is expected for this type of KT transition) at this point.
We obtain a rough estimate of $\varepsilon_{c_2}$, 
$\varepsilon_{c_2}^{c} \approx 1.65(15)$,
which is consistent with the results of our finite-size scaling
analysis, $\varepsilon_{c_2} =1.8(1)$ to within the accuracy of
the scaling.
Thus, the three estimates of the critical coupling,
$\varepsilon_{c_2}$, $\varepsilon_{c_2}^\beta$, and
$\varepsilon_{c_2}^c$ are consistent with one another.
Our best estimate is given by $\varepsilon_{c_2}$, since the other two
estimates are rougher and more likely to contain systematic errors.


%
\section{\label{sec:diss} Discussion}
We have analyzed the band-insulator-to-Mott-insulator transition
in the strong-coupling limit.
Using simple strong-coupling arguments,
we have derived an effective model
starting from the ionic Hubbard model.
The effective model, which we have formulated in a spin-one representation,
captures the physics of the transition and is less computationally
demanding than the ionic Hubbard model.
It contains spin-exchange processes which
are strongly restricted compared to those of a conventional spin
model.
In addition, the effective model demonstrates that
a single interaction parameter governs the transition.
Our density-matrix renormalization group study of this model confirms
that there are two transitions
at two clearly separated coupling strengths.
The system undergoes a transition 
from a band insulator to a spontaneously dimerized insulator
followed by a transition from the spontaneously dimerized phase to a
Mott insulator with increasing effective interaction.
This behavior corresponds to the behavior of the ionic Hubbard model
found in previous work.

\begin{figure}[htb]
  \centering
  \includegraphics[width=\mycolumnwidth,height=\brcolumnwidth]{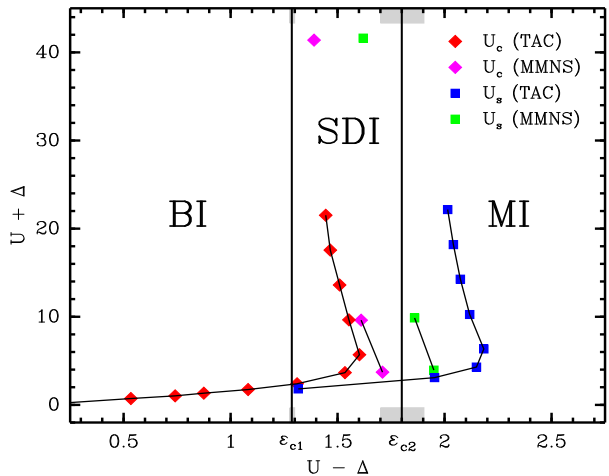}
  \caption{(Color online) Rotated ground-state phase diagram phase diagram 
    of the ionic Hubbard model depicting the phase boundaried obtained in
    Refs.~\onlinecite{Torio_2001} and {\onlinecite{Manmana_2004}}, 
    as well as the transition points $\varepsilon_{c_1} = 1.286(5)$ 
    and $\varepsilon_{c_2} = 1.8(1)$ obtained in this work, which
    apply in strong coupling in $U/t$, $\Delta/t$.
    The estimated error in our results are indicated by the gray-shaded
    bars at the upper and lower axes.}
  \label{fig:rotiHmPD}
\end{figure}

In Fig.~\ref{fig:rotiHmPD}, we explicitly compare the phase boundaries
obtained in our work  to phase boundaries obtained numerically for the
ionic Hubbard model in Refs.~\onlinecite{Torio_2001} and
\onlinecite{Manmana_2004}.
The phase diagram is plotted in the $45^\circ$ rotated $U$-$\Delta$
plane of the ionic Hubbard model, so that the abscissa corresponds to
our effective parameter $\varepsilon = U-\Delta$ and the intermediate
phase is expanded relative to the depiction in
Fig.~\ref{fig:iHmPD}. 
Since our effective model is based on a strong coupling expansion in
$U$ and $\Delta$, our results should be applicable to the ionic
Hubbard model in the large $U+\Delta$ limit.
As can be seen, for the BI-SDI boundary, both ionic Hubbard model
results tend towards our strong-coupling value as $U+\Delta$ becomes
larger, although the largest coupling point (at $\Delta=20$) from
Ref.~\onlinecite{Manmana_2004} is still outside our error bars.
The results for the SDI-MI transition boundary have larger
discrepancies, but our results lie between strong $U+\Delta$
extrapolations of the phase boundaries of
Ref.~\onlinecite{Manmana_2004} and that of
Ref.~\onlinecite{Torio_2001}.
This underlines the difficulty of obtaining the transition point in
this infinite-order Kosterlitz-Thouless transition.
To within large, but realistic error bars, the three sets of results
for this phase boundary are not necessarily inconsistent with each
other.
%

%
%
%

%
Our extraction of the critical exponents
for the first transition confirms 
that it belongs to the two-dimensional Ising universality class.
We have also shown that the universal scaling relations 
are fulfilled to within our numerical accuracy. 
At the second transition,
we have observed that
the mass gap closes exponentially
and that all relevant susceptibilities diverge.
An analysis of the scaling of 
the mass gap and of the bond-order susceptibility
confirms typical Kosterlitz-Thouless behavior.
The mass gap closes at the critical point 
and then remains zero as the interaction is further increased.
The susceptibility diverges 
in the entire critical region above the second transition point.
The overall scenario, with an Ising-like transition from the band
insulator to the spontaneously dimerized insulator followed by an
infinite-order transition from the dimerized insulator the Mott
insulator, is in complete agreement with the field-theoretical
prediction for the ionic Hubbard model.\cite{Fabrizio_1999}
The evolution of the appropriately mapped gaps with increasing
$\epsilon$ in the effective model is consistent with the picture
obtained for the ionic Hubbard model in Ref.~\onlinecite{Manmana_2004}.
Deep in the band insulating phase, all gaps in the spin and charge
sectors are equal and are set by the band gap.
As $\varepsilon_{c_1}$ is approached, the exciton gap,
defined as the energy gap between the ground state and
the first singlet excited state,  is the mass gap, and it goes to zero
at $\varepsilon_{c_1}$, while the spin gap (the gap to spin triplet
excitations) remains finite.
For $\varepsilon_{c_1} < \varepsilon < \varepsilon_{c_2}$
the mass gap is set by the gap to the
lowest-lying triplet, i.e., the spin gap, which is degenerate with
singlet excited states.
This gap goes to zero at $\varepsilon_{c_2}$.
For $\varepsilon > \varepsilon_{c_2}$, the spin and exciton gaps
remain zero, as expected in a critical phase, but the gaps
to add or remove one or more particles remain finite.
We note also that the mapping of electronic systems to spin-one systems
derived here can be adapted to a larger class of 
similar models or to generalizations of the ionic Hubbard model,
e.g., to chains with an ionic potential with a different periodicity
or even to two-dimensional systems.
Another potentially interesting application would be to relate exactly 
solvable spin-one models to electronic models and {\it vice versa}
via the spin-one composite representation.~\cite{Batista_2004}
%


%
\begin{acknowledgments}
L.~T.~would like to thank B.\ Normand, G.~Japaridze and C.~Hamer
for useful discussions and the DFG for
support through IRTG 790, ``Electron-Electron Interactions in Solids''.
This work was also supported by the Swiss National Foundation through the
National Center of Competence in Research ``Materials with Novel
Electronic Properties--MaNEP''.
\end{acknowledgments}
%


%
\appendix*
\section{\label{app:modder}Derivation of the effective model}
\begin{figure}[htbp]
  \centering
  \vspace{0.5cm}
  \includegraphics[angle=-90,width=\mycolumnwidth]{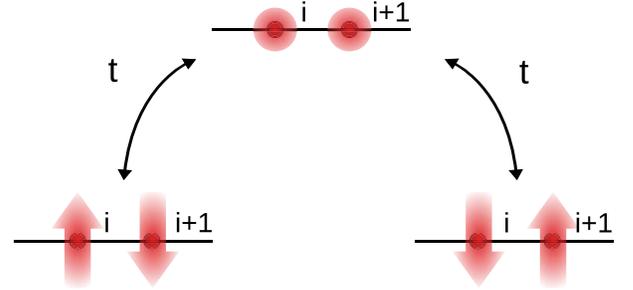}
  \caption{(Color online)
    Sketch of the allowed processes, 
    which are a relatively small subset of 
    those of the isotropic Heisenberg spin-chain model.}
  \label{fig:eff-hop}
\end{figure}

The effective Hamiltonian can most easily be derived by 
first  expressing the original Hamiltonian as a function of the
Hubbard operators
$\Xop_{i}^{\alpha \beta} = |\alpha_{i} \rangle \langle \beta_{i}|$,
where the $|\alpha_i\rangle$ and $|\beta_i\rangle$ designate an
element of the Hubbard basis 
$\{ |0\rangle , |\uparrow\rangle , |\downarrow\rangle , |d\rangle\}$ 
on site $i$.
Explicitly, they can be expressed as
\small
\begin{equation}
\Xop = 
\begin{bmatrix}
(\one-\nop_\dwr)(\one-\nop_\upr) & \cop_\upr(\one-\nop_\dwr) & \cop_\dwr (\one-\nop_\upr) & \cop_\dwr \cop_\upr\\
\cop_\upr^\dag (\one-\nop_\dwr) & (\one-\nop_\dwr)\nop_\upr & \cop_\upr^\dag \cop_\dwr & - \cop_\dwr \nop_\upr\\
\cop_\dwr^\dag (\one-\nop_\upr) & \cop_\dwr^\dag \cop_\upr & \nop_\dwr (\one-\nop_\upr) & \nop_\dwr \cop_\upr\\
\cop_\upr^\dag \cop_\dwr^\dag & - \cop_\dwr^\dag \nop_\upr & \nop_\dwr \cop_\upr^\dag & \nop_\dwr \nop_\upr
\end{bmatrix}
\, .
\nonumber
\end{equation}
\normalsize
For instance, we rewrite the ionic potential 
and the Coulomb interaction as
\begin{eqnarray}
\hat{H}_U & = & U \sum_{i=1}^{L} \Xop_i^{dd} \nonumber \\
          & = & U \sum_{j=1}^{L/2} \left( \Xop_{2j-1}^{dd} + \Xop_{2j}^{dd} \right)  \nonumber
\label{equ:coul-Xop-1}
\end{eqnarray}
and
\begin{eqnarray}
\hat{H}_\Delta & = & \frac{\Delta}{2} 
\sum_{i=1}^{L} \left(-1\right)^{i}
\left( \Xop_i^{\upr \upr} + \Xop_i^{\dwr \dwr} + 2 \Xop_i^{dd} \right) \nonumber \\ 
& = & \frac{\Delta}{2} 
\sum_{j=1}^{L/2} \left( -\Xop_{2j-1}^{\upr \upr} - \Xop_{2j-1}^{\dwr \dwr} - 2 \Xop_{2j-1}^{dd}
\right. \nonumber \\
& \quad & \qquad \left. + \Xop_{2j}^{\upr \upr} + \Xop_{2j}^{\dwr \dwr} + 2 \Xop_{2j}^{dd} \right) \nonumber \,.
\label{equ:coul-Xop-2}
\end{eqnarray}
They can then be mapped onto the spin-one model expressed in terms
of the operators $L_{i}^{s s'} = |s_i \rangle \langle s_i'|$, with 
$|s_i \rangle$ the spin-one $S_z$ basis $\{ | 1\rangle , | 0\rangle , | -1\rangle\}$
on site $i$. 
The single-site Hilbert space truncation is defined as
\begin{equation}
\left\{
\begin{array}{ll}
\Xop^{\alpha \beta}_{i} \rightarrow   0 & \text{for} \ \alpha \ \text{or} \ \beta = 0 \ \text{and} \ i = 2j-1 \\
\Xop^{\alpha \beta}_{i} \rightarrow   0 & \text{for} \ \alpha \ \text{or} \ \beta = d \ \text{and} \ i = 2j  \\
\Xop^{\alpha \beta}_{i} = \Lop^{\alpha \beta} & \text{otherwise.}
\end{array}
\right. 
\label{equ:trunc-op}
\end{equation}
In the spin-one basis,
\small
\begin{equation}
\Lop = 
\begin{bmatrix}
\frac{\left( \Sop_i^z \right)^2 + \Sop_i^z }{2} & \frac{\Sop_i^z\Sop_i^+}{\sqrt{2}} & \frac{\left( \Sop_i^+ \right)^2}{2} \\
\frac{\Sop_i^-\Sop_i^z}{\sqrt{2}} & \hat{1}_i-\left( \Sop_i^z \right)^2 & -\frac{\Sop_i^+\Sop_i^z}{\sqrt{2}} \\
\frac{\left( \Sop_i^- \right)^2}{2} & - \frac{\Sop_i^z\Sop_i^-}{\sqrt{2}} & \frac{\left( \Sop_i^z \right)^2 - \Sop_i^z }{2}
\end{bmatrix}
\, .
\label{equ:Loper}
\end{equation}
\normalsize
\newline
Hence, the interaction and the potential parts are transformed to
\begin{eqnarray}
\hat{H}_U  & = & U \sum_{j=1}^{L/2} \Lop_{2j-1}^{00}\, ,
\label{eq:int-Lop}
\end{eqnarray}
\begin{eqnarray}
\hat{H}_\Delta 
& = &
- \frac{\Delta}{2} 
\sum_{j=1}^{L/2} 
\left( 
\Lop_{2j-1}^{11} 
+ \Lop_{2j-1}^{-1-1}
+ 2 \Lop_{2j-1}^{00} 
\right. \nonumber \\
&\quad  & \qquad 
\left. 
- \Lop_{2j}^{11} 
- \Lop_{2j}^{-1-1} 
\right)
\,.
\label{equ:ioni-Lop}
\end{eqnarray}
Altogether, defining the coupling constant $\varepsilon = U - \Delta$,
the doping $\delta = N - L$, 
and writing the terms using spin-one operators, see Eq.~(\ref{equ:Loper}),
the two-term contribution becomes
\begin{equation}
\hat{H}^{e}_{\varepsilon} = 
- \frac{\varepsilon}{2} \sum_{i=1}^{L} \left( \Sop_i^z \right)^2 
- \frac{\varepsilon}{2} L - \frac{U}{2} \delta\, .
\label{eq:eff-int}
\end{equation}
Likewise, the hopping part is translated to
\begin{eqnarray}
\hat{H}^{e}_t  & = & t \sum_{i=1}^L
\left( \Lop^{0-1}_i \Lop^{01}_{i+1} -
\Lop^{01}_i \Lop^{0-1}_{i+1} 
\right. \nonumber \\
& & \qquad \left. 
+ \Lop^{-10}_i \Lop^{10}_{i+1} -
\Lop^{10}_i \Lop^{-10}_{i+1} \right)
\label{eq:hop-Lop}
\end{eqnarray}
or, in the spin one language
\begin{eqnarray}
\hat{H}^{e}_t  & = & \frac{t}{2} 
\sum_{i=1}^L \left( - \Sop^{+}_i\Sop^{z}_i  \Sop^{-}_{i+1}\Sop^{z}_{i+1}
\right. \nonumber \\
& & \qquad 
\left. 
+ \Sop^{-}_i\Sop^{z}_i  \Sop^{+}_{i+1}\Sop^{z}_{i+1} + \text{h.c.} \right)
\, ,
\label{eq:eff-hop-one}
\end{eqnarray}\\[0.2cm]
which is equivalent to Eq.~(\ref{equ:HeffectiveKin}).
A sketch of the allowed processes is shown in Fig.~\ref{fig:eff-hop}. 
These processes are a relative small subset of those of the isotropic 
Heisenberg spin chain model.
Note that the AFM exchange in the IHM maps 
to a sequence consisting of two scattering processes in the effective model.
%
%
%
%

%


%
\bibliography{biblio} 

\begin{thebibliography}{50}
\expandafter\ifx\csname natexlab\endcsname\relax\def\natexlab#1{#1}\fi
\expandafter\ifx\csname bibnamefont\endcsname\relax
  \def\bibnamefont#1{#1}\fi
\expandafter\ifx\csname bibfnamefont\endcsname\relax
  \def\bibfnamefont#1{#1}\fi
\expandafter\ifx\csname citenamefont\endcsname\relax
  \def\citenamefont#1{#1}\fi
\expandafter\ifx\csname url\endcsname\relax
  \def\url#1{\texttt{#1}}\fi
\expandafter\ifx\csname urlprefix\endcsname\relax\def\urlprefix{URL }\fi
\providecommand{\bibinfo}[2]{#2}
\providecommand{\eprint}[2][]{\url{#2}}

\bibitem[{\citenamefont{Nagaosa and
  Takimoto}(1986{\natexlab{a}})}]{Nagaosa_1986a}
\bibinfo{author}{\bibfnamefont{N.}~\bibnamefont{Nagaosa}} \bibnamefont{and}
  \bibinfo{author}{\bibfnamefont{J.}~\bibnamefont{Takimoto}},
  \bibinfo{journal}{J.~Phys.~Soc.~Jpn.} \textbf{\bibinfo{volume}{55}},
  \bibinfo{pages}{2735} (\bibinfo{year}{1986}{\natexlab{a}}).

\bibitem[{\citenamefont{Nagaosa and
  Takimoto}(1986{\natexlab{b}})}]{Nagaosa_1986b}
\bibinfo{author}{\bibfnamefont{N.}~\bibnamefont{Nagaosa}} \bibnamefont{and}
  \bibinfo{author}{\bibfnamefont{J.}~\bibnamefont{Takimoto}},
  \bibinfo{journal}{J.~Phys.~Soc.~Jpn.} \textbf{\bibinfo{volume}{55}},
  \bibinfo{pages}{2745} (\bibinfo{year}{1986}{\natexlab{b}}).

\bibitem[{\citenamefont{Nagaosa}(1986)}]{Nagaosa_1986c}
\bibinfo{author}{\bibfnamefont{N.}~\bibnamefont{Nagaosa}},
  \bibinfo{journal}{J.~Phys.~Soc.~Jpn.} \textbf{\bibinfo{volume}{55}},
  \bibinfo{pages}{2754} (\bibinfo{year}{1986}).

\bibitem[{\citenamefont{Torrance et~al.}(1981)\citenamefont{Torrance, Vazquez,
  Mayerle, and Lee}}]{Torrance_1981}
\bibinfo{author}{\bibfnamefont{J.~B.} \bibnamefont{Torrance}},
  \bibinfo{author}{\bibfnamefont{J.~E.} \bibnamefont{Vazquez}},
  \bibinfo{author}{\bibfnamefont{J.~J.} \bibnamefont{Mayerle}},
  \bibnamefont{and} \bibinfo{author}{\bibfnamefont{V.~Y.} \bibnamefont{Lee}},
  \bibinfo{journal}{Phys.~Rev.~Lett.} \textbf{\bibinfo{volume}{46}},
  \bibinfo{pages}{253} (\bibinfo{year}{1981}).

\bibitem[{\citenamefont{Horiuchi et~al.}(2003)\citenamefont{Horiuchi, Okimoto,
  Kumai, and Tokura}}]{Horiuchi_2003}
\bibinfo{author}{\bibfnamefont{S.}~\bibnamefont{Horiuchi}},
  \bibinfo{author}{\bibfnamefont{Y.}~\bibnamefont{Okimoto}},
  \bibinfo{author}{\bibfnamefont{R.}~\bibnamefont{Kumai}}, \bibnamefont{and}
  \bibinfo{author}{\bibfnamefont{Y.}~\bibnamefont{Tokura}},
  \bibinfo{journal}{Science} \textbf{\bibinfo{volume}{299}},
  \bibinfo{pages}{229} (\bibinfo{year}{2003}).

\bibitem[{\citenamefont{Gammel et~al.}(1992)\citenamefont{Gammel, Saxena,
  Batisti\'{c}, Bishop, and Phillpot}}]{Gammel_1992}
\bibinfo{author}{\bibfnamefont{J.~T.} \bibnamefont{Gammel}},
  \bibinfo{author}{\bibfnamefont{A.}~\bibnamefont{Saxena}},
  \bibinfo{author}{\bibfnamefont{I.}~\bibnamefont{Batisti\'{c}}},
  \bibinfo{author}{\bibfnamefont{A.~R.} \bibnamefont{Bishop}},
  \bibnamefont{and} \bibinfo{author}{\bibfnamefont{S.~R.}
  \bibnamefont{Phillpot}}, \bibinfo{journal}{Phys.~Rev.~B}
  \textbf{\bibinfo{volume}{45}}, \bibinfo{pages}{6408} (\bibinfo{year}{1992}).

\bibitem[{\citenamefont{Yamamoto}(2001)}]{Yamamoto_2001}
\bibinfo{author}{\bibfnamefont{S.}~\bibnamefont{Yamamoto}},
  \bibinfo{journal}{Phys.~Rev.~B} \textbf{\bibinfo{volume}{63}},
  \bibinfo{pages}{125124} (\bibinfo{year}{2001}).

\bibitem[{\citenamefont{Essler et~al.}(2005)\citenamefont{Essler, Frahm,
  G{\"o}hmann, Kl{\"u}mper, and Korepin}}]{Essler_2005}
\bibinfo{author}{\bibfnamefont{F.~H.~L.} \bibnamefont{Essler}},
  \bibinfo{author}{\bibfnamefont{H.}~\bibnamefont{Frahm}},
  \bibinfo{author}{\bibfnamefont{F.}~\bibnamefont{G{\"o}hmann}},
  \bibinfo{author}{\bibfnamefont{A.}~\bibnamefont{Kl{\"u}mper}},
  \bibnamefont{and} \bibinfo{author}{\bibfnamefont{V.~E.}
  \bibnamefont{Korepin}}, \emph{\bibinfo{title}{The One-Dimensional Hubbard
  Model}} (\bibinfo{publisher}{Cambridge University Press},
  \bibinfo{year}{2005}).

\bibitem[{\citenamefont{Torio et~al.}(2001)\citenamefont{Torio, Aligia, and
  Ceccatto}}]{Torio_2001}
\bibinfo{author}{\bibfnamefont{M.~E.} \bibnamefont{Torio}},
  \bibinfo{author}{\bibfnamefont{A.~A.} \bibnamefont{Aligia}},
  \bibnamefont{and} \bibinfo{author}{\bibfnamefont{H.~A.}
  \bibnamefont{Ceccatto}}, \bibinfo{journal}{Phys.~Rev.~B}
  \textbf{\bibinfo{volume}{64}}, \bibinfo{pages}{121105}
  (\bibinfo{year}{2001}).

\bibitem[{\citenamefont{Manmana et~al.}(2004)\citenamefont{Manmana, Meden,
  Noack, and Sch{\"o}nhammer}}]{Manmana_2004}
\bibinfo{author}{\bibfnamefont{S.~R.} \bibnamefont{Manmana}},
  \bibinfo{author}{\bibfnamefont{V.}~\bibnamefont{Meden}},
  \bibinfo{author}{\bibfnamefont{R.~M.} \bibnamefont{Noack}}, \bibnamefont{and}
  \bibinfo{author}{\bibfnamefont{K.}~\bibnamefont{Sch{\"o}nhammer}},
  \bibinfo{journal}{Phys.~Rev.~B} \textbf{\bibinfo{volume}{70}},
  \bibinfo{pages}{155115} (\bibinfo{year}{2004}).

\bibitem[{\citenamefont{Kampf et~al.}(2003)\citenamefont{Kampf, Sekania,
  Japaridze, and Brune}}]{Kampf_2003}
\bibinfo{author}{\bibfnamefont{A.~P.} \bibnamefont{Kampf}},
  \bibinfo{author}{\bibfnamefont{M.}~\bibnamefont{Sekania}},
  \bibinfo{author}{\bibfnamefont{G.~I.} \bibnamefont{Japaridze}},
  \bibnamefont{and} \bibinfo{author}{\bibfnamefont{P.}~\bibnamefont{Brune}},
  \bibinfo{journal}{J.~Phys.~: Condens.~Matter} \textbf{\bibinfo{volume}{15}},
  \bibinfo{pages}{5895} (\bibinfo{year}{2003}).

\bibitem[{\citenamefont{Kakashvili and Japaridze}(2004)}]{Kakashvili_2004}
\bibinfo{author}{\bibfnamefont{P.}~\bibnamefont{Kakashvili}} \bibnamefont{and}
  \bibinfo{author}{\bibfnamefont{G.}~\bibnamefont{Japaridze}},
  \bibinfo{journal}{J.~Phys.~C: Solid State Phys.}
  \textbf{\bibinfo{volume}{16}}, \bibinfo{pages}{5815} (\bibinfo{year}{2004}).

\bibitem[{\citenamefont{Resta and Sorella}(1995)}]{Resta_1995}
\bibinfo{author}{\bibfnamefont{R.}~\bibnamefont{Resta}} \bibnamefont{and}
  \bibinfo{author}{\bibfnamefont{S.}~\bibnamefont{Sorella}},
  \bibinfo{journal}{Phys.~Rev.~Lett.} \textbf{\bibinfo{volume}{74}},
  \bibinfo{pages}{4738} (\bibinfo{year}{1995}).

\bibitem[{\citenamefont{Fabrizio et~al.}(1999)\citenamefont{Fabrizio, Gogolin,
  and Nersesyan}}]{Fabrizio_1999}
\bibinfo{author}{\bibfnamefont{M.}~\bibnamefont{Fabrizio}},
  \bibinfo{author}{\bibfnamefont{A.~O.} \bibnamefont{Gogolin}},
  \bibnamefont{and} \bibinfo{author}{\bibfnamefont{A.~A.}
  \bibnamefont{Nersesyan}}, \bibinfo{journal}{Phys.~Rev.~Lett.}
  \textbf{\bibinfo{volume}{83}}, \bibinfo{pages}{2014} (\bibinfo{year}{1999}).

\bibitem[{\citenamefont{Delfino and Mussardo}(1998)}]{Delfino_1998}
\bibinfo{author}{\bibfnamefont{G.}~\bibnamefont{Delfino}} \bibnamefont{and}
  \bibinfo{author}{\bibfnamefont{G.}~\bibnamefont{Mussardo}},
  \bibinfo{journal}{Nucl.~Phys.~B} \textbf{\bibinfo{volume}{516}},
  \bibinfo{pages}{674} (\bibinfo{year}{1998}).

\bibitem[{\citenamefont{Fabrizio et~al.}(2000)\citenamefont{Fabrizio, Gogolin,
  and Nersesyan}}]{Fabrizio_2000}
\bibinfo{author}{\bibfnamefont{M.}~\bibnamefont{Fabrizio}},
  \bibinfo{author}{\bibfnamefont{A.~O.} \bibnamefont{Gogolin}},
  \bibnamefont{and} \bibinfo{author}{\bibfnamefont{A.~A.}
  \bibnamefont{Nersesyan}}, \bibinfo{journal}{Nucl.~Phys.~B}
  \textbf{\bibinfo{volume}{580}}, \bibinfo{pages}{647} (\bibinfo{year}{2000}).

\bibitem[{\citenamefont{Gidopoulos et~al.}(2000)\citenamefont{Gidopoulos,
  Sorella, and Tosatti}}]{Gidopoulos_2000}
\bibinfo{author}{\bibfnamefont{N.}~\bibnamefont{Gidopoulos}},
  \bibinfo{author}{\bibfnamefont{S.}~\bibnamefont{Sorella}}, \bibnamefont{and}
  \bibinfo{author}{\bibfnamefont{E.}~\bibnamefont{Tosatti}},
  \bibinfo{journal}{Eur.~Phys.~J.~B} \textbf{\bibinfo{volume}{14}},
  \bibinfo{pages}{217} (\bibinfo{year}{2000}).

\bibitem[{\citenamefont{Wilkens and Martin}(2001)}]{Wilkens_2001}
\bibinfo{author}{\bibfnamefont{T.}~\bibnamefont{Wilkens}} \bibnamefont{and}
  \bibinfo{author}{\bibfnamefont{R.~M.} \bibnamefont{Martin}},
  \bibinfo{journal}{Phys.~Rev.~B} \textbf{\bibinfo{volume}{63}},
  \bibinfo{pages}{235108} (\bibinfo{year}{2001}).

\bibitem[{\citenamefont{Lou et~al.}(2003)\citenamefont{Lou, Qin, Xiang, Chen,
  Tian, and Su}}]{Lou_2003}
\bibinfo{author}{\bibfnamefont{J.}~\bibnamefont{Lou}},
  \bibinfo{author}{\bibfnamefont{S.}~\bibnamefont{Qin}},
  \bibinfo{author}{\bibfnamefont{T.}~\bibnamefont{Xiang}},
  \bibinfo{author}{\bibfnamefont{C.}~\bibnamefont{Chen}},
  \bibinfo{author}{\bibfnamefont{G.-S.} \bibnamefont{Tian}}, \bibnamefont{and}
  \bibinfo{author}{\bibfnamefont{Z.}~\bibnamefont{Su}},
  \bibinfo{journal}{Phys.~Rev.~B} \textbf{\bibinfo{volume}{68}},
  \bibinfo{pages}{045110} (\bibinfo{year}{2003}).

\bibitem[{\citenamefont{Otsuka and Nakamura}(2005)}]{Otsuka_2005}
\bibinfo{author}{\bibfnamefont{H.}~\bibnamefont{Otsuka}} \bibnamefont{and}
  \bibinfo{author}{\bibfnamefont{M.}~\bibnamefont{Nakamura}},
  \bibinfo{journal}{Phys.~Rev.~B} \textbf{\bibinfo{volume}{71}},
  \bibinfo{pages}{155105} (\bibinfo{year}{2005}).

\bibitem[{\citenamefont{Legeza and S{\'o}lyom}(2006)}]{Legeza_2005}
\bibinfo{author}{\bibfnamefont{{\"O}.}~\bibnamefont{Legeza}} \bibnamefont{and}
  \bibinfo{author}{\bibfnamefont{J.}~\bibnamefont{S{\'o}lyom}},
  \bibinfo{journal}{Phys.~Rev.~Lett.} \textbf{\bibinfo{volume}{96}},
  \bibinfo{pages}{116401} (\bibinfo{year}{2006}).

\bibitem[{\citenamefont{Legeza et~al.}(2006)\citenamefont{Legeza, Buchta, and
  S{\'o}lyom}}]{Legeza_2006}
\bibinfo{author}{\bibfnamefont{{\"O}.}~\bibnamefont{Legeza}},
  \bibinfo{author}{\bibfnamefont{K.}~\bibnamefont{Buchta}}, \bibnamefont{and}
  \bibinfo{author}{\bibfnamefont{J.}~\bibnamefont{S{\'o}lyom}},
  \bibinfo{journal}{Phys.~Rev.~B} \textbf{\bibinfo{volume}{73}},
  \bibinfo{pages}{165124} (\bibinfo{year}{2006}).

\bibitem[{\citenamefont{Aligia and Batista}(2005)}]{Aligia_2005}
\bibinfo{author}{\bibfnamefont{A.~A.} \bibnamefont{Aligia}} \bibnamefont{and}
  \bibinfo{author}{\bibfnamefont{C.~D.} \bibnamefont{Batista}},
  \bibinfo{journal}{Phys.~Rev.~B} \textbf{\bibinfo{volume}{71}},
  \bibinfo{pages}{125110} (\bibinfo{year}{2005}).

\bibitem[{\citenamefont{Tincani}(2008)}]{Tincani_2008}
\bibinfo{author}{\bibfnamefont{L.}~\bibnamefont{Tincani}}, Ph.D. thesis,
  \bibinfo{school}{University of Marburg} (\bibinfo{year}{2008}).

\bibitem[{\citenamefont{Fazekas}(1999)}]{Fazekas_1999}
\bibinfo{author}{\bibfnamefont{P.}~\bibnamefont{Fazekas}},
  \emph{\bibinfo{title}{Lecture Notes on Electron Correlation and Magnetism}}
  (\bibinfo{publisher}{World Scientific}, \bibinfo{year}{1999}).

\bibitem[{\citenamefont{Haley and Erd{\"o}s}(1972)}]{Haley_1972}
\bibinfo{author}{\bibfnamefont{S.~B.} \bibnamefont{Haley}} \bibnamefont{and}
  \bibinfo{author}{\bibfnamefont{P.}~\bibnamefont{Erd{\"o}s}},
  \bibinfo{journal}{Phys.~Rev.~B} \textbf{\bibinfo{volume}{5}},
  \bibinfo{pages}{1106} (\bibinfo{year}{1972}).

\bibitem[{\citenamefont{Horovitz and S{\'o}lyom}(1987)}]{Horovitz_1987}
\bibinfo{author}{\bibfnamefont{B.}~\bibnamefont{Horovitz}} \bibnamefont{and}
  \bibinfo{author}{\bibfnamefont{J.}~\bibnamefont{S{\'o}lyom}},
  \bibinfo{journal}{Phys.~Rev.~B} \textbf{\bibinfo{volume}{35}},
  \bibinfo{pages}{7081} (\bibinfo{year}{1987}).

\bibitem[{\citenamefont{White}(1992)}]{White_1992}
\bibinfo{author}{\bibfnamefont{S.~R.} \bibnamefont{White}},
  \bibinfo{journal}{Phys.~Rev.~Lett.} \textbf{\bibinfo{volume}{69}},
  \bibinfo{pages}{2863} (\bibinfo{year}{1992}).

\bibitem[{\citenamefont{Schollw{\"o}ck}(2005)}]{Schollwock_2005}
\bibinfo{author}{\bibfnamefont{U.}~\bibnamefont{Schollw{\"o}ck}},
  \bibinfo{journal}{Rev.~Mod.~Phys.} \textbf{\bibinfo{volume}{77}},
  \bibinfo{pages}{259} (\bibinfo{year}{2005}).

\bibitem[{\citenamefont{Fabrizio and Gogolin}(1995)}]{Fabrizio_1995}
\bibinfo{author}{\bibfnamefont{M.}~\bibnamefont{Fabrizio}} \bibnamefont{and}
  \bibinfo{author}{\bibfnamefont{A.~O.} \bibnamefont{Gogolin}},
  \bibinfo{journal}{Phys.~Rev.~B} \textbf{\bibinfo{volume}{51}},
  \bibinfo{pages}{17827} (\bibinfo{year}{1995}).

\bibitem[{\citenamefont{Bed{\"u}rftig et~al.}(1998)\citenamefont{Bed{\"u}rftig,
  Brendel, Frahm, and Noack}}]{Bedurftig_1998}
\bibinfo{author}{\bibfnamefont{G.}~\bibnamefont{Bed{\"u}rftig}},
  \bibinfo{author}{\bibfnamefont{B.}~\bibnamefont{Brendel}},
  \bibinfo{author}{\bibfnamefont{H.}~\bibnamefont{Frahm}}, \bibnamefont{and}
  \bibinfo{author}{\bibfnamefont{R.~M.} \bibnamefont{Noack}},
  \bibinfo{journal}{Phys.~Rev.~B} \textbf{\bibinfo{volume}{58}},
  \bibinfo{pages}{10225} (\bibinfo{year}{1998}).

\bibitem[{\citenamefont{Hotta and Shibata}(2006)}]{Hotta_2006}
\bibinfo{author}{\bibfnamefont{C.}~\bibnamefont{Hotta}} \bibnamefont{and}
  \bibinfo{author}{\bibfnamefont{N.}~\bibnamefont{Shibata}},
  \bibinfo{journal}{Physica B} \textbf{\bibinfo{volume}{378}},
  \bibinfo{pages}{1039} (\bibinfo{year}{2006}).

\bibitem[{\citenamefont{Legeza and F{\'a}th}(1996)}]{Legeza_1996}
\bibinfo{author}{\bibfnamefont{{\"O}.}~\bibnamefont{Legeza}} \bibnamefont{and}
  \bibinfo{author}{\bibfnamefont{G.}~\bibnamefont{F{\'a}th}},
  \bibinfo{journal}{Phys.~Rev.~B} \textbf{\bibinfo{volume}{53}},
  \bibinfo{pages}{14349} (\bibinfo{year}{1996}).

\bibitem[{\citenamefont{Legeza et~al.}(2003)\citenamefont{Legeza, R{\"o}der,
  and Hess}}]{Legeza_2003b}
\bibinfo{author}{\bibfnamefont{{\"O}.}~\bibnamefont{Legeza}},
  \bibinfo{author}{\bibfnamefont{J.}~\bibnamefont{R{\"o}der}},
  \bibnamefont{and} \bibinfo{author}{\bibfnamefont{B.~A.} \bibnamefont{Hess}},
  \bibinfo{journal}{Phys.~Rev.~B} \textbf{\bibinfo{volume}{67}},
  \bibinfo{pages}{125114} (\bibinfo{year}{2003}).

\bibitem[{\citenamefont{Legeza and S{\'o}lyom}(2003)}]{Legeza_2003}
\bibinfo{author}{\bibfnamefont{{\"O}.}~\bibnamefont{Legeza}} \bibnamefont{and}
  \bibinfo{author}{\bibfnamefont{J.}~\bibnamefont{S{\'o}lyom}},
  \bibinfo{journal}{Phys.~Rev.~B} \textbf{\bibinfo{volume}{68}},
  \bibinfo{pages}{195116} (\bibinfo{year}{2003}).

\bibitem[{\citenamefont{Kogut}(1979)}]{Kogut_1979}
\bibinfo{author}{\bibfnamefont{J.~B.} \bibnamefont{Kogut}},
  \bibinfo{journal}{Rev.~Mod.~Phys.} \textbf{\bibinfo{volume}{51}},
  \bibinfo{pages}{659} (\bibinfo{year}{1979}).

\bibitem[{\citenamefont{Barber}(1983)}]{Barber_1983}
\bibinfo{author}{\bibfnamefont{M.~N.} \bibnamefont{Barber}}, in
  \emph{\bibinfo{booktitle}{Phase Transition and Critical Phenomena}}, edited
  by \bibinfo{editor}{\bibfnamefont{C.}~\bibnamefont{Domb}} \bibnamefont{and}
  \bibinfo{editor}{\bibfnamefont{M.~S.} \bibnamefont{Green}}
  (\bibinfo{publisher}{Academic Press}, \bibinfo{year}{1983}),
  vol.~\bibinfo{volume}{8} of \emph{\bibinfo{series}{Phase Transition and
  Critical Phenomena}}, chap.~\bibinfo{chapter}{2}.

\bibitem[{\citenamefont{Glaus and Schneider}(1984)}]{Glaus_1984}
\bibinfo{author}{\bibfnamefont{U.}~\bibnamefont{Glaus}} \bibnamefont{and}
  \bibinfo{author}{\bibfnamefont{T.}~\bibnamefont{Schneider}},
  \bibinfo{journal}{Phys.~Rev.~B} \textbf{\bibinfo{volume}{30}},
  \bibinfo{pages}{215} (\bibinfo{year}{1984}).

\bibitem[{\citenamefont{Hamer and Barber}(1980)}]{Hamer_1980}
\bibinfo{author}{\bibfnamefont{C.~J.} \bibnamefont{Hamer}} \bibnamefont{and}
  \bibinfo{author}{\bibfnamefont{M.~N.} \bibnamefont{Barber}},
  \bibinfo{journal}{J.~Phys.~A: Math.~Gen.} \textbf{\bibinfo{volume}{13}},
  \bibinfo{pages}{L169} (\bibinfo{year}{1980}).

\bibitem[{\citenamefont{Hamer and Barber}(1981{\natexlab{a}})}]{Hamer_1981}
\bibinfo{author}{\bibfnamefont{C.~J.} \bibnamefont{Hamer}} \bibnamefont{and}
  \bibinfo{author}{\bibfnamefont{M.~N.} \bibnamefont{Barber}},
  \bibinfo{journal}{J.~Phys.~A: Math.~Gen.} \textbf{\bibinfo{volume}{14}},
  \bibinfo{pages}{241} (\bibinfo{year}{1981}{\natexlab{a}}).

\bibitem[{\citenamefont{Hamer and Barber}(1981{\natexlab{b}})}]{Hamer_1981b}
\bibinfo{author}{\bibfnamefont{C.~J.} \bibnamefont{Hamer}} \bibnamefont{and}
  \bibinfo{author}{\bibfnamefont{M.~N.} \bibnamefont{Barber}},
  \bibinfo{journal}{J.~Phys.~A: Math.~Gen.} \textbf{\bibinfo{volume}{14}},
  \bibinfo{pages}{259} (\bibinfo{year}{1981}{\natexlab{b}}).

\bibitem[{\citenamefont{Hamer}(1983)}]{Hamer_1983}
\bibinfo{author}{\bibfnamefont{C.~J.} \bibnamefont{Hamer}},
  \bibinfo{journal}{J.~Phys.~A: Math.~Gen.} \textbf{\bibinfo{volume}{16}},
  \bibinfo{pages}{3085} (\bibinfo{year}{1983}).

\bibitem[{\citenamefont{Zhu et~al.}(2003)\citenamefont{Zhu, Garst, Rosch, and
  Si}}]{Zhu_2003}
\bibinfo{author}{\bibfnamefont{L.}~\bibnamefont{Zhu}},
  \bibinfo{author}{\bibfnamefont{M.}~\bibnamefont{Garst}},
  \bibinfo{author}{\bibfnamefont{A.}~\bibnamefont{Rosch}}, \bibnamefont{and}
  \bibinfo{author}{\bibfnamefont{Q.}~\bibnamefont{Si}},
  \bibinfo{journal}{Phys.~Rev.~Lett.} \textbf{\bibinfo{volume}{91}},
  \bibinfo{pages}{066404} (\bibinfo{year}{2003}).

\bibitem[{\citenamefont{Feynman}(1939)}]{Feynman_1939}
\bibinfo{author}{\bibfnamefont{R.~P.} \bibnamefont{Feynman}},
  \bibinfo{journal}{Phys.~Rev.} \textbf{\bibinfo{volume}{56}},
  \bibinfo{pages}{340} (\bibinfo{year}{1939}).

\bibitem[{\citenamefont{Press et~al.}(1999)\citenamefont{Press, Flannery,
  Teukolsky, and Vetterling}}]{Press_1999}
\bibinfo{author}{\bibfnamefont{W.~H.} \bibnamefont{Press}},
  \bibinfo{author}{\bibfnamefont{B.~P.} \bibnamefont{Flannery}},
  \bibinfo{author}{\bibfnamefont{S.~A.} \bibnamefont{Teukolsky}},
  \bibnamefont{and} \bibinfo{author}{\bibfnamefont{W.~T.}
  \bibnamefont{Vetterling}}, \emph{\bibinfo{title}{Numerical Recipes in C++,
  The Art of Scientific Computing}} (\bibinfo{publisher}{Cambridge University
  Press}, \bibinfo{year}{1999}).

\bibitem[{\citenamefont{{Degli Esposti Boschi} and
  Ortolani}(2004)}]{Boschi_2004}
\bibinfo{author}{\bibfnamefont{C.}~\bibnamefont{{Degli Esposti Boschi}}}
  \bibnamefont{and} \bibinfo{author}{\bibfnamefont{F.}~\bibnamefont{Ortolani}},
  \bibinfo{journal}{Eur.~Phys.~J.~B} \textbf{\bibinfo{volume}{41}},
  \bibinfo{pages}{503} (\bibinfo{year}{2004}).

\bibitem[{\citenamefont{{Di Francesco} et~al.}(1999)\citenamefont{{Di
  Francesco}, Mathieu, and S{\'e}n{\'e}chal}}]{DiFrancesco_1999}
\bibinfo{author}{\bibfnamefont{P.}~\bibnamefont{{Di Francesco}}},
  \bibinfo{author}{\bibfnamefont{P.}~\bibnamefont{Mathieu}}, \bibnamefont{and}
  \bibinfo{author}{\bibfnamefont{D.}~\bibnamefont{S{\'e}n{\'e}chal}},
  \emph{\bibinfo{title}{Conformal Field Theory}}
  (\bibinfo{publisher}{Springer}, \bibinfo{year}{1999}).

\bibitem[{\citenamefont{Calabrese and Cardy}(2004)}]{Calabrese_2004}
\bibinfo{author}{\bibfnamefont{P.}~\bibnamefont{Calabrese}} \bibnamefont{and}
  \bibinfo{author}{\bibfnamefont{J.}~\bibnamefont{Cardy}},
  \bibinfo{journal}{J.~Stat.~Mech.: Theor.~Exp.}
  \textbf{\bibinfo{volume}{2004}}, \bibinfo{pages}{P06002}
  (\bibinfo{year}{2004}).

\bibitem[{\citenamefont{Kosterlitz}(1974)}]{Kosterlitz_1974}
\bibinfo{author}{\bibfnamefont{J.~M.} \bibnamefont{Kosterlitz}},
  \bibinfo{journal}{J.~Phys.~C: Solid State Phys.}
  \textbf{\bibinfo{volume}{7}}, \bibinfo{pages}{1046} (\bibinfo{year}{1974}).

\bibitem[{\citenamefont{Batista and Ortiz}(2004)}]{Batista_2004}
\bibinfo{author}{\bibfnamefont{C.~D.} \bibnamefont{Batista}} \bibnamefont{and}
  \bibinfo{author}{\bibfnamefont{G.}~\bibnamefont{Ortiz}},
  \bibinfo{journal}{Adv.~Phys.} \textbf{\bibinfo{volume}{53}},
  \bibinfo{pages}{1} (\bibinfo{year}{2004}).

\end{thebibliography}
%


%
\end{document}